\documentclass[review]{llncs}

\usepackage{lineno,hyperref}
\usepackage{booktabs} 
\usepackage{tikz}
\usepackage{pgfplots}
\usepackage{pgfplotstable}
\usepackage{graphicx}
\usetikzlibrary{pgfplots.statistics}
\usetikzlibrary{arrows,shapes,positioning,shadows,trees}
\usepackage{colortbl}
\usepackage{subfig}
\usepackage{pifont}
\usepackage{tikzsymbols}
\usepackage{multirow}
\usepackage{ulem}
\usepackage{url}
\usepackage{lipsum}
\makeatletter
\def\ps@pprintTitle{%
 \let\@oddhead\@empty
 \let\@evenhead\@empty
 \def\@oddfoot{\centerline{\thepage}}%
 \let\@evenfoot\@oddfoot}
\makeatother

\begin{document}

\begin{frontmatter}

\title{IntelliAV: Building an Effective On-Device Android Malware Detector}

\author{Mansour Ahmadi\inst{1,2}\thanks{Corresponding author: m.ahmadi@northeastern.edu}, Angelo Sotgiu\inst{1}, and Giorgio Giacinto\inst{1}}

\institute{University of Cagliari, Italy
\and
Northeastern University, USA}




\maketitle

\begin{abstract}
The importance of employing machine learning for malware detection has become explicit to the security community. 
Several anti-malware vendors have claimed and advertised the application of machine learning in their products in which the inference phase is performed on servers and high-performance machines, but the feasibility of such approaches on mobile devices with limited computational resources has not yet been assessed by the research community, vendors still being skeptical.
In this paper, we aim to show the practicality of devising a learning-based anti-malware on Android mobile devices, first. Furthermore, we aim to demonstrate the significance of such a tool to cease new and evasive malware that can not easily be caught by signature-based or offline learning-based security tools. To this end, we first propose the extraction of a set of lightweight yet powerful features from Android applications. Then, we embed these features in a vector space to build an effective as well as efficient model. Hence, the model can perform the inference on the device for detecting potentially harmful applications. We show that without resorting to any signatures and relying only on a training phase involving a reasonable set of samples, the proposed system, named \texttt{IntelliAV} \footnote{This paper is the extended version of \texttt{IntelliAV} conference paper \cite{10.1007/978-3-319-66808-6_10}}\footnote{ IntelliAV application is available online: http://www.intelliav.com}, provides more satisfying performances than the popular major anti-malware products. Moreover, we evaluate the robustness of \texttt{IntelliAV} against common obfuscation techniques where most of the anti-malware solutions get affected.
\end{abstract}


\end{frontmatter}


\section{Introduction}

Android is the most popular platform for mobile devices, with almost 85\% of the market share in the first quarter of 2017 \cite{IDC:2017:Online}. 
More interestingly, Android is now the most popular operating system connected to the Internet after overtaking Windows \cite{statcounter}. 
The majority of the security issues affecting Android systems can be attributed to third-party applications (app) rather than to the Android OS itself.
According to a report in 2017 from G DATA (a security vendor), a new instance of Android malware emerges nearly every 10 seconds \cite{GDATA}.
Besides, another recent report from McAfee shows that the malware infection rate of Android mobile devices is soaring \cite{mcafee}. We believe that this huge amount of mobile malware needs to be timely detected, possibly by intelligent tools running on the device, because it has been shown that malware can bypass offline security checks (e.g., by relying on so-called droppers, that load the malicious payload after being activated), and live in the wild for a while. In fact, to the best of our knowledge, even the most recent versions of Android anti-malware products are still not satisfactory to cope with most of the novel and obfuscated malware. 

Today, machine learning is one of the most successful helper techniques for Android malware detection and classification \cite{STREAM,Drebin,DroidMiner,DroidSieve}. The recent advances in the optimization of machine learning tools that can execute on mobile platforms, such as Android \cite{tensorflowmobile}, increase the possibility of empowering security applications with machine learning tools. Despite the improvement in processor and RAM of mobile devices, the development of any mobile anti-malware product should take into account the processing time to avoid battery drain, in particular when machine learning techniques are employed, as they are known to be computational demanding. On the other hand, we observe that a learning-based Android anti-malware product does not need to be necessarily sophisticated, as it has been shown that Android malware seems to perform simpler tasks than the desktop counterparts \cite{aresu2015}. All the reasons above stimulate a proposal for an on-device machine learning solution to detect potential malicious applications.

\textbf{Contribution.} Accordingly, in this paper, we introduce an intelligent risk-based anti-malware approach for Android devices, called \texttt{IntelliAV}, which is built on top of the open-source and multi-platform TensorFlow library.
In fact, we illustrate the feasibility and the advantages of such an approach on the device by leveraging on the existing literature, and, in particular, on previous works by the authors, to tackle the deficiencies of existing Android anti-malware products, mostly based on pattern matching techniques, as well as offline learning-based approaches.
As far as we know, the performances of learning-based malware detection systems for Android have been only tested off-device, i.e., with the availability of computational power and memory space well beyond the capabilities of mobile devices.
More specifically, the three main contributions of \texttt{IntelliAV} are as follows:
\begin{itemize}
  \item[$(i)$] We design a system relying on a trained model on a sizeable set of applications. The model is carefully constructed to be both effective and efficient by wisely selecting a set of lightweight, discriminative, and effective features. Moreover, the model is precisely validated
by tuning its parameters to be practical for the capabilities of Android devices. We then show how the crafted model can be embedded in the \texttt{IntelliAV} application, and can readily be deployed on Android devices.
  \item[$(ii)$] The performances of \texttt{IntelliAV} are evaluated through a cross-validation process, where our system can achieve 92\% detection rate, that is comparable to other off-device learning-based Android malware detection while relying on a comparatively small set of features. Moreover, as a supplementary experiment, \texttt{IntelliAV} is analyzed on two different sets of more recent malware samples with respect to the samples included in the training specimens. Interestingly, \texttt{IntelliAV} can achieve 96\% detection rate from an independent test by a 3rd party organization, and it obtains 72\% detection rate ---on a set of randomly gathered apps by us--- that is higher than the performances of the top 5 commercial Android anti-malware products.
  \item[$(iii)$] To understand the robustness of \texttt{IntelliAV}, we evaluate the impact of two common evasion techniques, i.e., dropper and obfuscation, on the proposed system. For droppers, interestingly,  we show how \texttt{IntelliAV} can stop them on the device while offline machine learning techniques would fail to detect. For the second examination, we prove the power of \texttt{IntelliAV} on identifying heavily obfuscated malware, which perfectly shows how machine learning can add to the cost evasion.
\end{itemize}

\textbf{Paper organization.} The rest of the paper is organized as follows:\\
First, we highlight the importance of an on-device risk-based malware detection (\S\ref{sec:motivation})
and review the existing works on this area (\S\ref{sect:related}).
Next, we reveal the structure of \texttt{IntelliAV}, motivating the choice of features,
and the procedure followed to construct the model (\S\ref{sect:sys}). 
We then assess the approach (\S\ref{sect:exp}) and remark 
the limitations of \texttt{IntelliAV} (\S\ref{sec:limitations}). 
Finally, we conclude our paper discussing future directions of \texttt{IntelliAV} (\S\ref{sect:conclusions}).
\section{Motivation of having a Risk-based On-Device Approach}
\label{sec:motivation}

There are several incentives for the security community to develop a risk-based mobile security approach, possibly based on machine learning, as well as performing real-time on the mobile device in addition to employing it on servers.

\subsection{Malware in Google Play store}
There have been various reports on the practicability of by-passing Google security mechanisms by malware coders. Consequently, malware keeps sneaking on the Google Play store and remains accessible to users until a security vendor/researcher reports it to Google as harmful. For instance, the Check Point security firm described a zero-day mobile ransomware found on Google Play in January 2017 \cite{checkpoint-gp}. This malware was dubbed as a \textit{Charger} application and more than a million users downloaded this app. Another report from the same vendor indicates the case of new variants of the famous Android malware family HummingBad \cite{checkpoint-gp1} on Google Play. 
Another specimen of malware that could infiltrate Google Play is a packed malware that sends fraudulent premium SMS messages and charges users for fake services without their knowledge \cite{jsmalware}.
More detail on vetting these samples is available in Section~\ref{sec:comparison}.

\subsection{Install Malware from unknown sources}
Third-party app stores are popular among mobile users because they usually offer applications at great discounts, as well as users from specific countries can find there those applications whose access is restricted by the Google Play store according to each country's rules. Nevertheless, security checks on the third-party stores are not as effective as those available in the Google Play store. Therefore, third-party markets are a breeding ground for mobile malware propagation, and this fact sometimes leads people to download spoofed versions of well-known applications. A large number of reports on malicious applications found in these stores have been published during the past few years. In addition to the third-party markets, direct download from unknown websites is another source of infection. It is quite often that users can be eluded by fake tempting titles like free games when browsing the web, so that applications are downloaded and installed directly on devices from untrusted websites. Another source of contamination is though phishing SMS messages that contain links to malicious applications. Recent reports by Lookout and Google \cite{pegasus-lookout,pegasus-google} show how a targeted attack malware, namely \textit{Chrysaor}, which is presumed of infecting devices via a phishing attack, could remain undetected for a few years. More detail on vetting these samples is available in Section~\ref{sec:comparison}.

\subsection{Distribute malware in supply chain}
Users of mobile devices are frequently recommended by information security experts to be cautious when downloading applications from untrusted sources or even when they install not very popular apps from Google Play. However, there are fewer warnings for the users on the reliability of safety claims of new devices, especially when they might be shipped with pre-installed malware. Despite the concerns of manufacturers on securing the `supply chain', it can be compromised by attackers, for the number of people and companies involved in the supply chain of the components.
There is a recent report that shows how devious hackers spread malware on Android devices somewhere along the supply chain before the user obtained the phone \cite{checkpoint}. More detail on vetting these samples is available in Section~\ref{sec:comparison}.

\subsection{Easy to evade pattern matching}
Almost all of the major Android anti-malware kernels operate, to the extent of our knowledge, by matching signatures or patterns. These types of scanning engines let both malware variants of known families, as well as zero-day malware threaten our own devices. There are claims by a few Android anti-malware vendors on the use of heuristic approaches like machine learning in their products. However, no evidence regarding the implementation of the kernel including machine learning on the device is available. Hence, to find out further, we evaluate them on very recent malware samples, as well as on a few obfuscated malware samples. More detail is available in Section~\ref{sec:comparison} and Section~\ref{sec:obfuscation}.

\subsection{Droppers dodge offline vetting tools}
A typical kind of evasion technique employed by any malware to evade off-device vetting mechanisms is by developing droppers, and Android malware makes no exception \cite{lookout-dropers}. Droppers don't directly perform malicious activities, and they are designed to install some sort of malicious application to a device. Therefore, detecting droppers is not a straightforward task even by advanced off-device machine learning techniques as a dropper itself usually exhibits a few standard behaviors that are common in legitimate applications as well, so that it does not obviously reveal malicious actions. We analyze a few representative malware samples using this attack vector in Section~\ref{sec:droppers}.

\subsection{Conclusion}
All of the above observations encourage to empower Android devices with a machine-learning anti-malware engine, either as a complement to pattern matching techniques, or as an independent complete solution.
\section{Related Work} \label{sect:related}

The problem of detecting Android malware through machine learning approaches has been explored quite a lot since 2010 \cite{Shabtai:2010}. While an entire overview is outside of the scope of this paper, we suggest the interested reader resorting to one of the recent surveys on this subject, e.g., the taxonomy proposed in \cite{androidsurvey}. Additionally, it is out of the scope of the paper a review of dynamic malware analysis approaches \cite{DroidScribe,STREAM,Crowdroid} as dynamic analysis has its specific advantages and pitfalls. For instance, we are dealing with an on-device tool, and it is not officially possible that a process accesses system calls of another process without root privileges, which makes the dynamic analysis approaches almost impractical on the end user device. Hence, we provide here some of the more closely relevant papers that rely on static analysis technique. The existing methods are classified into two distinct levels, namely off-device and on-device malware detection.

\subsection{Off-Device Malware Detection}
Offline testing usually has no hard computational restrictions, thanks to the availability of computational power compared to the one available on mobile devices. Some of the prominent malware detection models are MudFlow \cite{MudFlow}, AppAudit \cite{appaudit}, and DroidSIFT \cite{DroidSIFT} relying on information flow analysis\cite{FlowDroid}, while DroidMiner \cite{DroidMiner}, and MaMaDroid \cite{MaMaDroid} are based on API sequences. Although this allows constructing complex models capable of detecting malware with a very high accuracy, the use of elaborate features such as information flows and API sequences
makes these approach harder to be carried out on the device.
Lighter approaches, such as Drebin \cite{Drebin}, DroidAPIMiner \cite{DroidAPIMiner}, and DroidSieve \cite{DroidSieve} that make use of meta-data, as well as syntactic features, allow for their porting to on-device applications.

\subsection{On-Device Malware Detection}
Based on the best of our knowledge, there are a few approaches in the research community that used machine learning for on-device malware detection, and none of them is publicly available for performance comparison. Drebin \cite{Drebin} is one of them, which has been cited the most on this topic. While the paper shows some screenshots of the UI, the application itself is not available. Besides, while the proposed system is for both workstation and mobile devices, the actual needs for a learning-based malware detection engine on the device was not specified.
Among the commercial Android anti-malware tools, two of them claim to use machine learning techniques, as reported in Section~\ref{sec:comparison}, but the extent to which machine learning is used in these tools is not disclosed. Finally, Qualcomm recently announced the development of a machine learning tool for on-device mobile phone security, but the details of the system, as well as its performances, are not publicly available \cite{qualcomm}.

\begin{table}[!htb]
\centering
\scalebox{1}{
\begin{tabular}{clcccl}
\hline\noalign{\smallskip} 
\multirow{2}{*}{Year} & \multirow{2}{*}{Method} & \multicolumn{2}{c}{Detection} & \multirow{2}{*}{Feature} \\
& & \textbf{On-Device} & \textbf{Available} \\
\noalign{\smallskip}\hline\noalign{\smallskip}
2014 & DroidAPIMiner \cite{DroidAPIMiner} & $-$ & $-$ &  API,PKG,PAR \\
2014 & DroidMiner \cite{DroidMiner} & $-$ & $-$ &  CG,API SEQ  \\
2014 & Drebin \cite{Drebin} & \ding{51} & $-$ & PER,STR,API,INT \\ 
2014 & DroidSIFT \cite{DroidSIFT} & $-$ & $-$ &  API-F \\
2015 & AppAudit \cite{appaudit} & $-$ & \ding{51} & API-F \\
2015 & MudFlow \cite{MudFlow} & $-$ & \ding{51} &  API-F \\
2017 & MaMaDroid~\cite{MaMaDroid} & $-$ & \ding{51} & CG,API SEQ \\
2017 & DroidSieve~\cite{DroidSieve} & $-$ & $-$ & API,PER,INT,PN,STR,ST \\
2017 & Qualcomm ~\cite{qualcomm} & \ding{51} & $-$ & Not Available \\
\hline\noalign{\smallskip}
Ours & \textbf{IntelliAV} & \ding{51} & \ding{51} & PER,INT,API,ST \\
\noalign{\smallskip}\hline
\noalign{\smallskip}
\end{tabular}
}
\begin{flushleft}
{\scriptsize \texttt{API}: Application Programming Interface, \texttt{API-F}:
Information Flow between APIs, \texttt{INT}: Intents,  \texttt{CG}: Call Graph,
\texttt{SEQ}: Sequence,
\texttt{PER}: Requested Permissions, 
\texttt{STR}: Embedded strings,
\texttt{ST}: Statistical features, \texttt{PN}: Package names}
\end{flushleft}
\smallskip
\caption{Android Malware detection techniques based on machine learning techniques and static analysis. All of the systems that are mostly based on API, API-F, and API SEQ would fail against reflection. IntelliAV is the only on-device system that is available in the market.}
\label{tab:RelWorks}
\end{table}

\subsection{Summary}
As an overall comparison (see Table \ref{tab:RelWorks}) with the previous approaches, we believe that \texttt{IntelliAV} 
provides for an effective and practical on-device anti-malware solution for Android systems, totally based on machine learning techniques. \texttt{IntelliAV} is available online, and can move a step toward having an advanced security tool on mobile devices.
\section{System Design} \label{sect:sys}
Figure~\ref{fig:system-overview} illustrates the architecture of the proposed \texttt{IntelliAV} system.
Its design consists of the following two main phases:

\tikzstyle{block} = [rectangle, draw, fill=gray!20, 
    text width=5em, text centered, rounded corners, minimum height=4em]
    \tikzstyle{recblock} = [rectangle, draw, fill=gray!40, 
    text width=5em, text centered, minimum height=2em]
    \tikzstyle{subblock} = [rectangle, draw, fill=gray!9, 
    text width=5.5em, text centered, rounded corners, minimum height=2em]
\tikzstyle{dashedblock} = [rectangle, draw, dashed, fill=gray!9, 
    text width=5.5em, text centered, rounded corners, minimum height=2em]
\tikzstyle{line} = [draw, -latex']
    \tikzstyle{io} = [rectangle, 
    text width=5em, text centered, rounded corners, minimum height=2em]
        \tikzstyle{nio} = [rectangle,
    text width=5em, text centered, rounded corners, minimum height=1em]
        \tikzstyle{nblock} = [rectangle, draw, fill=gray!20,
    text width=5em, text centered, rounded corners, minimum height=4em]
    \tikzstyle{sblock} = [rectangle, draw, fill=gray!20,
    text width=5em, text centered, rounded corners, minimum height=2em]
\tikzstyle{ndashedblock} = [rectangle, draw, dashed, fill=white!9, 
    text width=5.5em, text centered, rounded corners, minimum height=2em]
\tikzstyle{horizontal-dotted} = [dash pattern = on 1pt off 4pt,
      line width =1.4pt]

\begin{figure}[t]
\resizebox{!}{6cm}{
\begin{tikzpicture}[node distance = 2cm, auto]
    \node [block, minimum height=6em, yshift=-1em, xshift=-2cm,text depth = 2.7cm, text width = 2.7cm] (flowdroid) {Feature\\Extraction};
    \node[dashedblock, text width = 2.3cm] at ([yshift=0em]flowdroid.center) (GCM) {Meta-Data};
    \node[dashedblock, text width = 2.3cm] at ([yshift=-2.4em]flowdroid.center) (GCM) {Dex Code};
    \node[inner sep=0pt, left of=flowdroid, xshift=-1cm] (dataset)
    {
    \includegraphics[width=.06\textwidth]{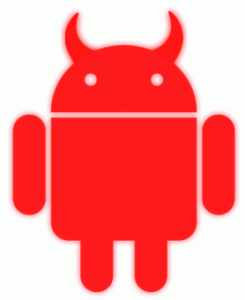}
    \includegraphics[width=.06\textwidth]{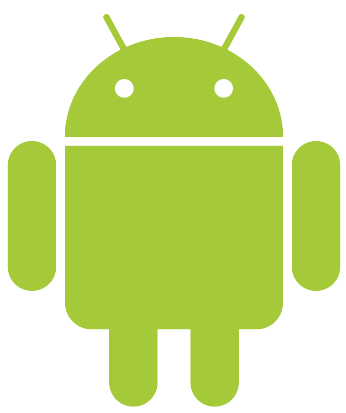}
    };
    \node [io, below of=dataset, yshift=0.6cm, text width = 3cm, xshift=-0.2em] (gcmper) {Training Apps};
    \node [block, right of=flowdroid, minimum height=9em, xshift=6cm, text depth = 2.2cm, text width = 6.5cm] (clustering) {Model Construction};
    \node[recblock, text width = 6.501cm, xshift=0em] at ([yshift=-4em]clustering.center) (tensorflow) {TensorFlow};
    \node[dashedblock, text width = 2.7cm, xshift=-1.7cm] at ([yshift=1em]clustering.center) (classifier1) {Model Selection};
    \node[dashedblock, right of=classifier1, text width = 2.7cm, xshift=-0.4cm] at ([yshift=1em]clustering.center) (classifier2) {Feature Selection};
    \node[dashedblock, text width = 2.7cm, xshift=-1.7cm] at ([yshift=-1.5em]clustering.center) (classifier3) {Validation};
    \node[dashedblock, right of=classifier3, text width = 2.9cm, xshift=-0.4cm] at ([yshift=-1.5em]clustering.center) (classifier4) {Parameter Tuning};
    \node[inner sep=0pt, below of=clustering, yshift=-12em, xshift=-2cm] (mc2)
    {
    \includegraphics[width=1.0\textwidth]{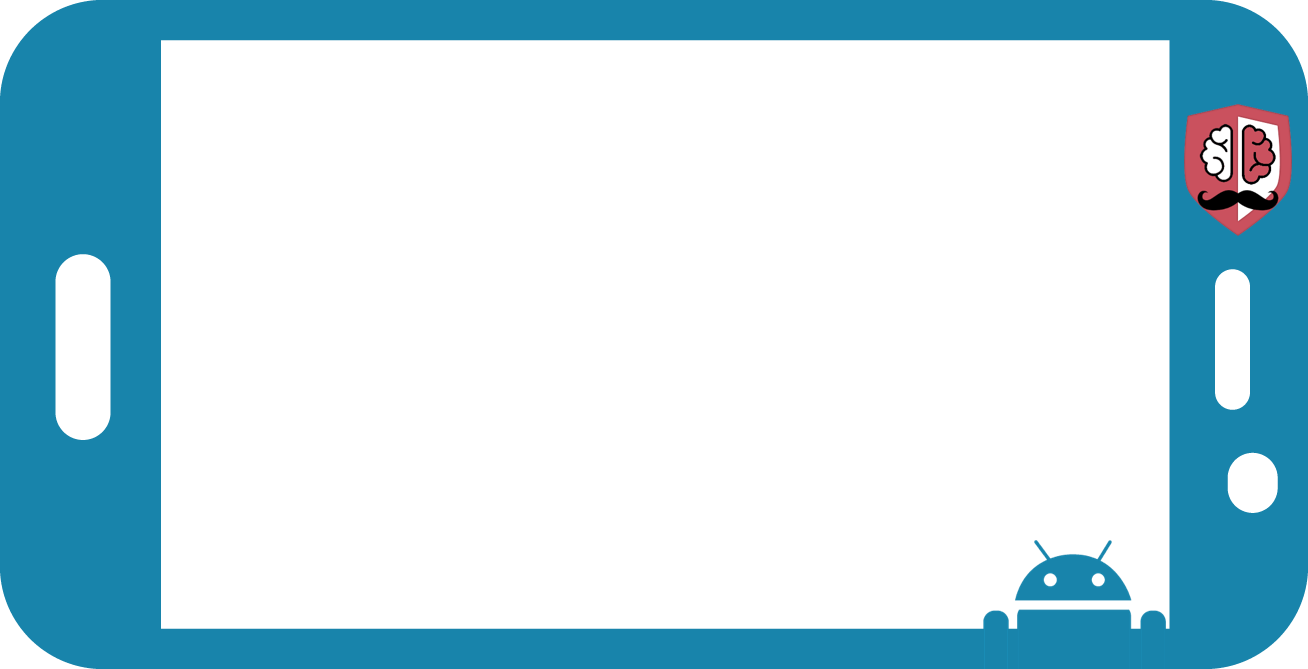}
    };
\node [block, minimum height=6em, xshift=7em, text depth = 2.7cm, text width = 2.7cm] at ([yshift=2.3em]mc2.center)  (devfeatextr) {Feature\\Extraction};
    \node[dashedblock, text width = 2.3cm] at ([yshift=0em]devfeatextr.center) (GCM) {Meta-Data};
    \node[dashedblock, text width = 2.3cm] at ([yshift=-2.4em]devfeatextr.center) (GCM) {Dex Code};    
    \node[inner sep=0pt, left of=mc2, yshift=-1em, xshift=0cm] at ([yshift=-2.5em]mc2.center) (test)
    {
    \includegraphics[width=.05\textwidth]{Images/mal-android.png}
    \includegraphics[width=.05\textwidth]{Images/ben-android.png}
    };
    \node [io, yshift=3em, below of=test, text width = 2cm, xshift=0em] (testcom) {Risk Score};
    \node [io, yshift=-2.5em, below of=devfeatextr, text width = 0cm, xshift=0em] (instapp) {App};
     \node [block, above of=test,minimum height=6em, xshift=0em, yshift=0em, text depth = 2.3cm, text width = 2.9cm] (prediction) {Prediction};
    \node[dashedblock, text width = 2.6cm] at ([yshift=-0.5em]prediction.center) (tflib) {TensorFlow Shared Objects Library (Native)};
     \draw[horizontal-dotted, below of=clustering, yshift=-5em, xshift=-2cm](-5,1)--(15,1);
     \node [io, yshift=0em, right of=clustering, text width = 3cm, xshift=14em, rotate=270] (ows) {\uline{{\large On Workstation}}};
     \node [io, yshift=0em, right of=mc2, text width = 3cm, xshift=20em, rotate=270] (omd) {\uline{{\large On Mobile Device}}};
    \path [line] (flowdroid) -- node[xshift=0cm, yshift=0em, text width=3cm, align=center] {Feature Vector}(clustering);
    \path [line] (devfeatextr) -- node[xshift=0cm, yshift=0em, text width=2cm, align=center] {Feature Vector}(prediction);
    \path [line,dashed] (dataset) -- (flowdroid);
    \path [line,dashed] (prediction) -- (test);
    \path [line,dashed] (instapp) -- (devfeatextr);
    \path [line] (clustering.south west) -| node[xshift=-1.9cm, yshift=-0.8em, text width=1cm, align=center] {Optimized Model}(prediction.north);

\end{tikzpicture}
}
\caption{Overview of \texttt{IntelliAV}.}\label{fig:system-overview}
\end{figure} 

\begin{enumerate}

\item[(\textit{i})] \textbf{Training the model offline.}
As a first step, we resort to a conventional computing environment to build a classification model.
To conduct the training phase, we gathered a relatively large number of applications (\S\ref{sec:data}). Then, a carefully selected set of characteristics (features) is extracted from the applications to learn a discriminant function allowing the distinction between malicious and benign behaviors (\S\ref{sec:features}).
Next, a classification function is learned by associating each feature vector to the 
type of applications it has been extracted from, i.e., malware or goodware (\S\ref{sec:classification}).

\item[(\textit{ii})] \textbf{Model operation on the device.}
As the second phase, the optimized model is embedded in the \texttt{IntelliAV} Android application so that \texttt{IntelliAV} can produce a risk score for each downloaded APK or installed apps on the device, without resorting to any external interaction, e.g., with cloud services (\S\ref{sec:test}).

\end{enumerate}

\subsection{Feature Engineering}
\label{sec:features}

The feature extraction step is the core phase for any learning-based system. To address Android malware detection, the security community has suggested various types of features as discussed in Section\ref{sect:related}. However, some sets of features related to primary Android functions, like permissions, APIs, and Intents, usually allow achieving reasonable detection results, with the aim to alert for the presence of probably harmful applications \cite{Drebin,DroidSieve}. Extracting this set of features is also feasible on mobile devices because they do not need deep static analysis, thus expecting a limited computational effort.

With the aim of extracting a set of efficient and effective features for our proposed system, we resort to the following four categories of characteristics. Three of them are derived from the `\textit{manifest}' of Android applications, namely Permissions,  Intent Filters, components statistics, and the fourth one is extracted from the DEX code, notably APIs.
A typical Android app has a single DEX file, and most of the standard malware detection approaches focus on this file.
Because of the Dalvik Executable specification, the total number of methods that can be referenced within a single DEX file is limited to 65,536 -including Android framework methods, library methods, and methods in the code. This limitation can be handled by multidex configuration if an application extends to more than 65K methods. Therefore, multidex can be a desirable technique for Android malware to split the payload into multiple DEX files to make the detection harder \cite{Symantec5evasion}. For this reason, to be more comprehensive in the terms of extracting the characteristics of apps, \texttt{IntelliAV} has to parse all of the DEX files of an application although it makes the feature extraction process somewhat slower (see section~\ref{sec:overhead}).

To construct the feature vector, we consider all the permissions and the intent-filters inquired by the samples included in the training set. Besides, four statistical features from application's components such as the total number of activities, services, broadcast receivers, and content providers are added to the feature vector as they can reveal somewhat the extent of abilities each application has. For instance, the number of activities in many malware categories is usually fewer than the number of activities available in goodware, except for the case of malware that is built by repackaging benign applications. 
Moreover, we manually select a set of 179 APIs as features and include them in the feature vector. The selected APIs are those that reveal some particular characteristics of the application that are known to be peculiar to either goodware or malware. For instance, the \texttt{invoke} API from the \texttt{java.lang.reflect.Method} class shows whether an application uses reflection (i.e., a technique for hiding APIs) or not. Note that permissions and APIs are coded as binary features, which means that their value is either one or zero depending on the feature being or not present in the application. If we considered the number of permissions, we would have ended up with useless information, as each permission needs to be declared just once in the \texttt{manifest}. The same reasoning motivates the use of binary feature to represent API usage. The main reason is that although it is possible to get the count of the usage of an API in an application, the procedure would increase the processing time without producing more useful information so that we ignored it. By contrast, intent-filters are integer-valued features, as they represent the number of times an intent-filter is declared in the manifest. Considering this count for intent-filter features makes them more meaningful rather than simply considering their presence or not in the application. Similarly,
the application's components are represented as integer-valued features, as we count the number of components for each different type (e.g., activities, services, etc.).

In total, the feature vector contains 4000 features. To avoid overfitting and make \texttt{IntelliAV} faster on the mobile device, we decided to reduce the number of features by selecting the most discriminative ones
through a feature selection procedure (see Section~\ref{sec:classification}). During the feature selection step, we consider four thresholds, namely, 25\%, 50\%, and 75\% of the top features, as well as the whole feature set. Interestingly, a model containing 25\% (i.e., 1000) of the top features can achieve the best result while uses fewer features. Therefore, the final assortment consists of 322 features related to permissions, 503 features linked to Intent filters, four statistical features from components (e.g., count of activities), and 171 features associated with API usage (see Table~\ref{tab:features}).

\begin{table}[t]
\centering
\caption{Features used in \texttt{IntelliAV}.}\label{tab:features}
\begin{tabular}{c|cc}
\hline
\rowcolor{gray!60}
Category & Number of Features & Type \\ \hline
\rowcolor{gray!30}
\multicolumn{3}{c}{\textbf{Meta-Data}} \\
Permissions & 322 & Binary \\
Intent Filters & 503 & Count \\
Statistical & 4 & Count \\
\rowcolor{gray!30}
\multicolumn{3}{c}{\textbf{Dex Code}} \\
APIs & 171 & Binary \\
\hline
\end{tabular}
\end{table}

\subsection{Model Construction}
\label{sec:classification}

As we mentioned earlier, the model needs to be trained on a workstation, and it is not necessary to conduct the training phase on the device because it has to be performed once we need to update the model according to the evolution of malware. 
In other words, in contrast to pattern matching techniques, our system does not require frequent updates by the end user, which diminishes the amount of computation and network traffic transferred.
The number of times the model needs to be updated should be quite small, as reports showed that just the 4\% of the total number of Android malware is actually from new malware families \cite{AV-TEST}. 

To discriminate malware from goodware, we need to rely on binary classification techniques. Over the past years, a large number of classification techniques have been proposed by the scientific community, and the choice of the most suitable classifier for a given task is often guided by preceding experience in different domains, as well as by trial-and-error procedures. However, among all of the existing classifiers, Random Forest classifier \cite{RandomForests} have shown high performances in a variety of tasks \cite{JMLR:14}. The Random Forests algorithm belongs to the ensemble learning methods in which many decision trees are constructed at training time by randomly selecting the features used by each decision tree. Eventually, the algorithm outputs the class of an instance at the testing time based on the combined decision of the tress. One of the main reasons that Random forest models
often achieve better results compared to others is that it is an ensemble classifier, which means it helps to reduce the variance in performances of the decision trees. So, the final model exhibits low bias and low variance, which makes the model more robust against both the \textit{underfitting} and \textit{overfitting} problems \cite{bishop:prml:book:2007}.

We build \texttt{IntelliAV} on the top of TensorFlow \cite{TensorFlow} library to be able to train our model offline, as well as to test it on Android devices. 
TensorFlow is an open source library for machine learning, which was published by Google in November 2015. 
A TensorFlow model is highly portable as it supports the vast majority of platforms such as Linux, Mac OS, Windows, and mobile computing platforms like Android and iOS. TensorFlow computations are expressed as data flow graphs. Nodes in the graph represent mathematical operations, while the graph edges represent the multidimensional data arrays (tensors) communicating between them.
To the best of our knowledge, \texttt{IntelliAV} is the first anti-malware tool that has proposed employing TensorFlow. 
More specifically, we employ an implementation of Random Forests in TensorFlow, called TensorForest \cite{TensorForest}. 

To simplify the learning task and reduce the risk of the so-called \textit{overfitting} problem, i.e., to avoid that the model fits the training set but exhibits a low generalization capability with respect to new unknown specimens, we employ feature selection to reduce the size of the feature set by excluding irrelevant and noisy features.
In particular, as done in \cite{Ahmadi:2016:NFE}, we computed the so-called \textit{mean decrease impurity} score for each feature, and retained those features which have been assigned the highest scores. Note that the mean decrease impurity technique is often referred to as the Gini impurity, or information gain criterion.

\subsection{On-Device Testing}
\label{sec:test}
As we mentioned before, TensorFlow facilitates the task of using machine learning models on mobile devices. So, we embed the trained model, obtained according to the procedure described in Section~\ref{sec:classification}, in \texttt{IntelliAV}. It is important to acknowledge that as far as the model is resident on the device and not in the cloud, the network traffic generated by \texttt{IntelliAV} is zero, compared to approaches that communicate with the cloud by sending a hash/APK to obtain the risk associated to an app.

The size of TensorFlow models depends on the complexity of the model. For instance, if the number of trees in the TensorForest model increases, consequently the size of the model increases as well. The size of \texttt{IntelliAV} model that we transferred to the device is about 14.1MB. Having said that, when it is embedded into the APK and because the model is originally saved in a textual format, the final size of the model after compression in the APK becomes just 3.3MB.

Whenever an application needs to be tested, first, \texttt{IntelliAV} loads the model, then extracts the features from the application on the device, and finally, it feeds the model by the extracted features to acquire the application's risk type.
The model provides a likelihood value between 0 and 1, denoting the degree of maliciousness of the application. Then, the likelihood is thresholded to three types of risks to make it more understandable for the end user.
We empirically provide the following guideline for interpreting the likelihood. If the likelihood is lower than 0.4, the risk is low and we suggest the user consider the application as safe. If the likelihood is between 0.4 and 0.6, then the application should be removed if the application is not so popular in Google Play or if the user is not sure about the trustworthiness of the source application. Finally, the application has to be removed with high confidence if the likelihood is higher than 0.6. These thresholds have been set after testing the system on a set containing different applications. It is worth to mention that the analyses of apps that identify potential harmful apps may yield both false positives and false negatives.

\begin{figure}[t]
   \centering
   ~~~~~~~
	\subfloat[Scan installed applications]{\includegraphics[width=0.35\textwidth]{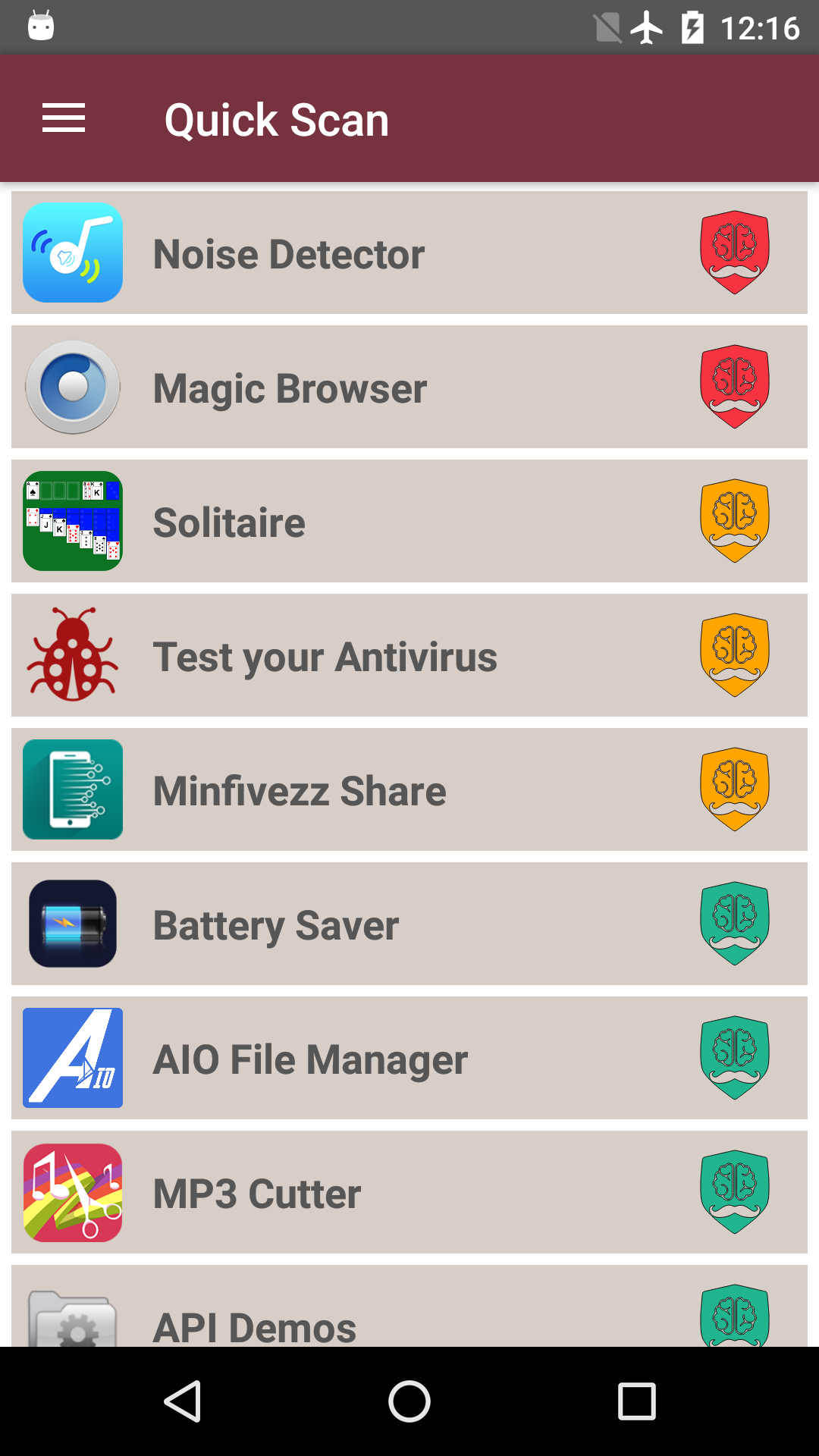}\label{fig:quickscan}}
	\hfill
    \subfloat[Scan an APK]{\includegraphics[width=0.35\textwidth]{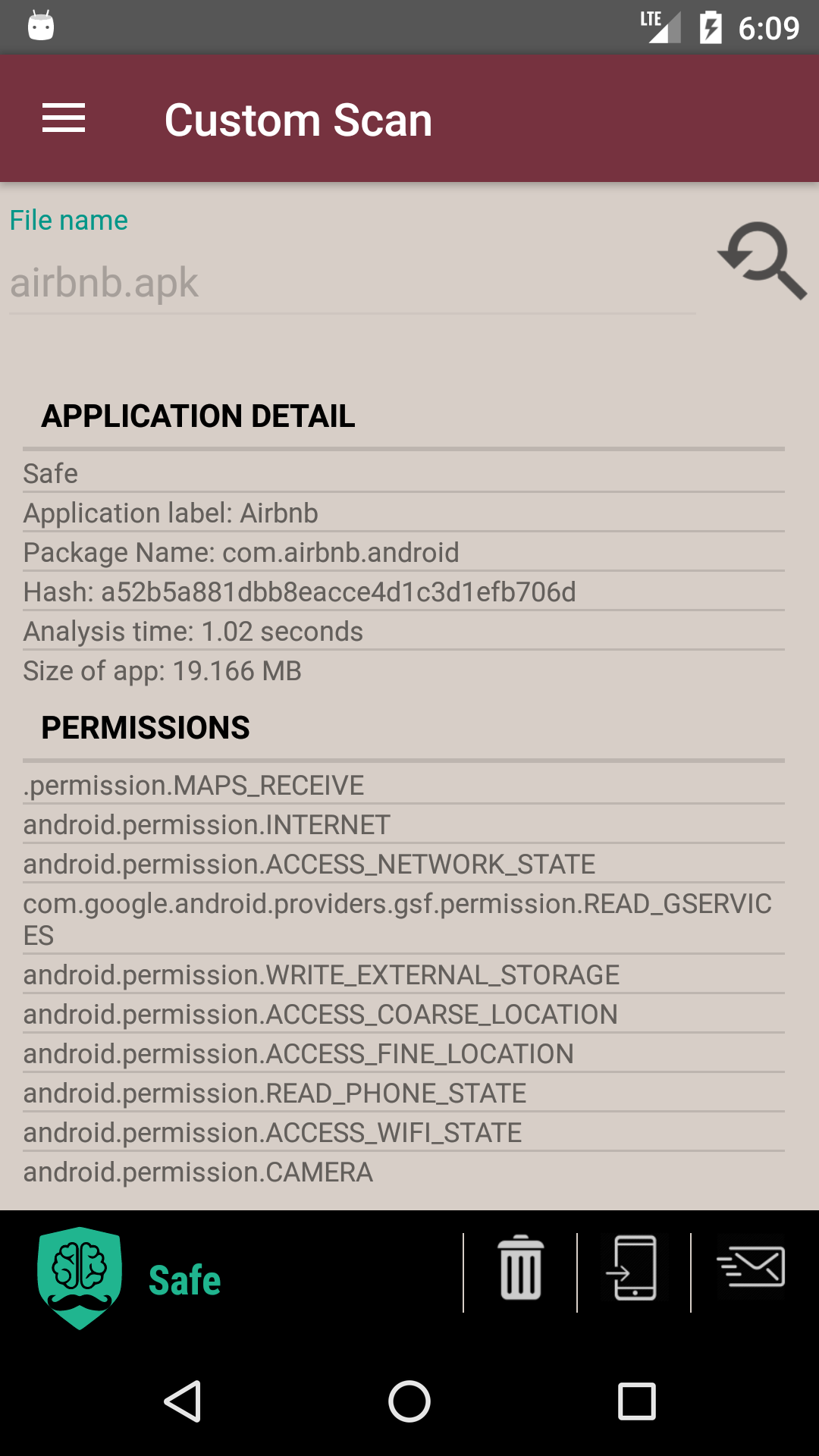}\label{fig:customscan}}
    \caption{\texttt{IntelliAV} abilities.}\label{fig:intelliavability}
\end{figure}

As shown in figure~\ref{fig:intelliavability}, the two main capabilities of \texttt{IntelliAV} are the verification of the risk of all the installed applications on the device (\textit{Quick Scan}), as well as analyzing the risk of downloaded/dropped APKs (\textit{Custom Scan}). The latter ability is necessary as it helps the user to check the risk of the application before installation. One of the prominent use cases of this ability is stopping droppers on the device, while offline machine learning systems can be simply lured by the use of droppers in malware (see Section~\ref{sec:droppers}).
For the \textit{Custom Scan}, \texttt{IntelliAV} needs the \texttt{READ\_EXTERNAL\_STORAGE} permission to access the contents of the application's package on the external storage. For the \textit{Quick Scan}, \texttt{IntelliAV} has to read \texttt{base.apk} file in a sub-directory with a name corresponding to the package name, which is located in the \textit{/data/app/} directory. As far as the permission of \texttt{base.apk} file is \texttt{-rw-r--r--}, which means every user can read the content of this file, \texttt{IntelliAV} needs neither any permission nor a rooted device to evaluate the installed applications.

\section{Experimental Analysis} \label{sect:exp}

In this section, we address the following research questions: 
\begin{itemize}
\item Is \texttt{IntelliAV} able to detect new and unseen malware (\S\ref{sec:unseenmalware} and \S\ref{sec:AV-TEST})?
\item Are the performances of \texttt{IntelliAV} comparable to the ones of popular mobile anti-malware products, although \texttt{IntelliAV} is completely based on machine learning techniques (\S\ref{sec:comparison})?
\item How much is \texttt{IntelliAV} robust against prevalent evasion techniques like droppers (\S\ref{sec:droppers}) and obfuscation (\S\ref{sec:obfuscation})?
\item Which is the overhead of \texttt{IntelliAV} on real devices (\S\ref{sec:overhead})?
\end{itemize}

Before addressing these questions, we explain the experimental settings and the data used for building our model (\S\ref{sec:data}).

\subsection{Experimental Setup}
\label{sec:data}
To train \texttt{IntelliAV}, we have collected 19,722 applications from VirusTotal\cite{virustotal}, divided into 10,058 benign and 9,664 malicious applications. 
We considered the diversity of malicious applications, by including samples belonging to different categories, such as Adware, Ransomware \cite{Heldroid}, and GCM malware \cite{Ahmadi:2016:gcm}.
In addition, in order to cover the variety of malware characteristics throughout its evolution during time, we also take into account samples from the first versions of Android malware, namely January 2011, until recent versions as of December 2016. We consider an application as being malicious if it was tagged as being malware by at least 10 of the tools used by VirusTotal. 

The whole process of feature extraction and model construction was carried out on a laptop with a 2 GHz quad-core processor and 8GB of memory. The two metrics that have been used for assessing the performance of our approach were the False Positive Rate (FPR) and the True Positive Rate (TPR).
FPR is the percentage of goodware samples misclassified as badware, while TPR is the fraction of correctly-detected badware samples (also known as detection rate). 
A Receiver-Operating-Characteristic (ROC) curve reports TPR against FPR for all possible model's decision thresholds, which can help to find the best decision threshold by considering the trade-off.

\subsection{Results}
\label{sec:result}
To adequately evaluate the effectiveness of \texttt{IntelliAV}, the following scenarios were considered.

\subsubsection{Cross Validation}
\label{sec:cv}
\begin{figure}[t]
\centering
\includegraphics[scale=0.4]{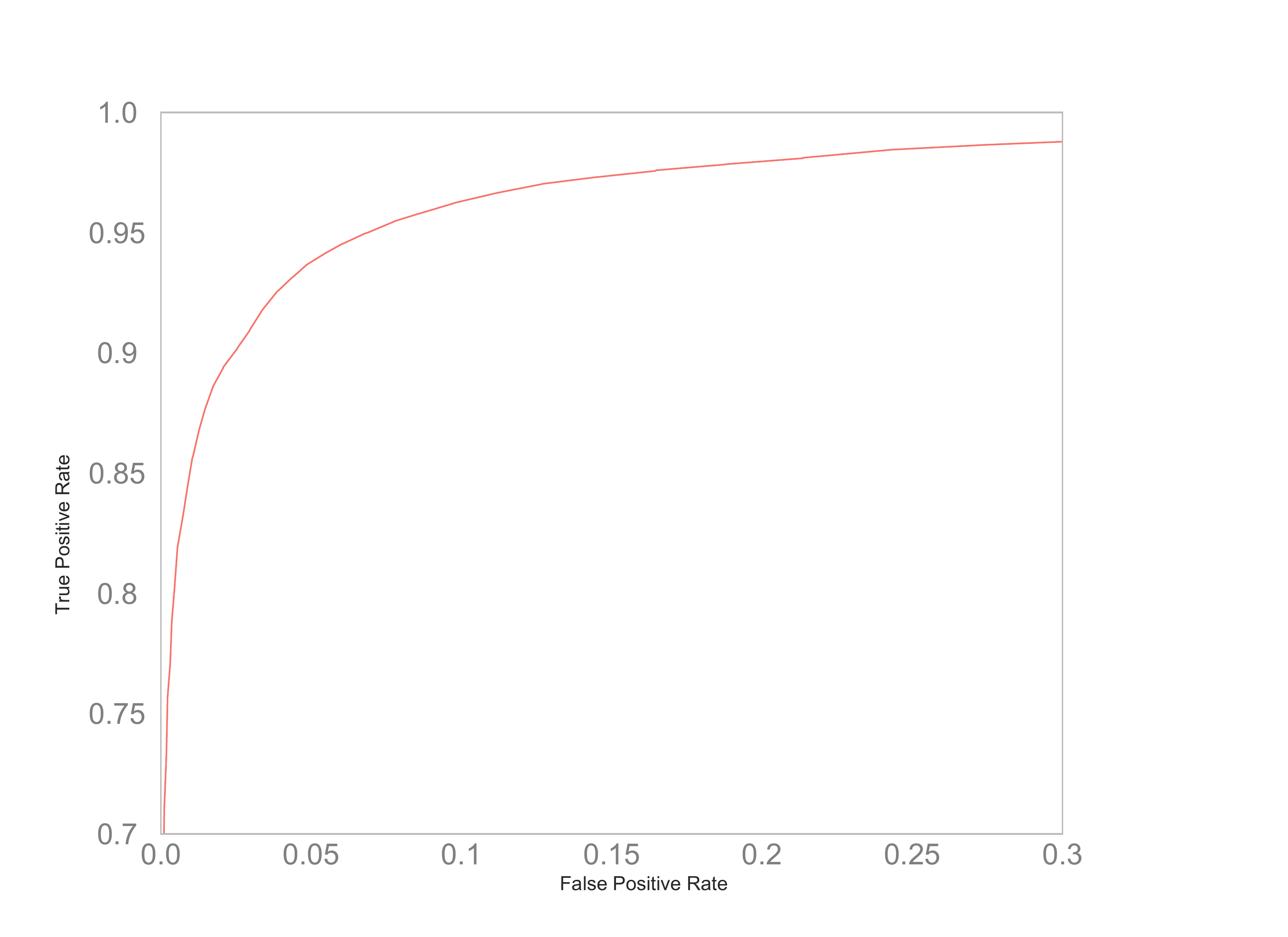}
\caption{ROC curve of TensorForest (5-fold cross validation). FPR and TPR are respectively 4.2\% and 92.5\%.}
\label{fig:roc}
\end{figure}

One might fit a model on the training set very well so that the model will perfectly classify all of the samples that are used during the training phase. However, this might not provide the model with the generalization capability, and that's why we evaluated the model by a cross-validation procedure to find the optimum-tuned parameters to be used for building the final model as a trade-off between correct detection and generalization capability.
Consequently, we evaluate \texttt{IntelliAV} on the set of applications described in Section~\ref{sec:data} through a 5-fold cross-validation procedure, to provide statistically-sound results. 
In this validation technique, samples are divided into 5 groups, called folds, with almost equal sizes.
The prediction model is created using 4 folds, and then it is tested on the final remaining fold. 
The procedure is repeated 5 times on different folds to be sure that each data point is evaluated exactly once. 
We repeat the procedure by running the Random Forest algorithm multiple times to obtain the most appropriate parameters while keeping the size of the model lower.
The ROC of the best-fitted model is shown in Figure~\ref{fig:roc}.
The values of FPR and TPR are respectively 4.2\% and 92.5\% which is quite acceptable although the set of considered features is relatively small, namely 1000 features.

\subsubsection{Evaluation on the training set}
\label{sec:signature}
Although cross-validation is important for setting the model parameters, we do not desire the model exhibit high misclassification on the training samples.
Therefore, as an additional investigation to verify the effectiveness of the tuned parameters based on the cross-validation procedure explained in Section~\ref{sec:cv}, we test the model, which was constructed on all 19.7K samples, on the same set, namely on all the samples used for training. 
Table~\ref{tab:sig} shows the classification results on the training set. It shows that \texttt{IntelliAV} misclassifies just a few training samples. This confirms how the model is carefully fitted on the training set, so that is able to correctly classify almost all of the training samples with very high accuracy, while it avoids being overfitted.

\begin{table}[t]
\centering
\caption{Training on the set of samples explained in Section~\ref{sec:data} and testing on the same set. GT refers to the Ground-truth of samples.}\label{tab:sig}
\begin{tabular}{c|c|cc}
\noalign{\smallskip}
\rowcolor{gray!60}
Train & \multicolumn{3}{c}{Test}\\
\multirow{2}{*}{\#Samples} & \multirow{2}{*}{GT (\#Samples)} & \multicolumn{2}{c}{Classified as}\\
 &  &  Malicious & Benign \\ \hline
\multirow{4}{*}{19,722} & Malicious (9,664) & 9,640 & 24 \\
 & & (TPR = 99.75\%) \\
& Benign (10,058) & 7 & 10,051 \\
 & & (FPR = 0.07\%) \\
\hline
\end{tabular}
\end{table}

\subsubsection{Evaluation on new Malware}
\label{sec:unseenmalware}

To realize the sustainability \cite{203684} of our model during a period of 3 months after the model is deployed, we examine the performance system on a set made up of 2311 malware samples, and 2898 benign applications, that have been first seen by VirusTotal between January and March of 2017. We consider an application as being malicious when it was labeled as malware by at least 5 of the tools used by VirusTotal. 
This set of test samples contains randomly selected applications that were newer than the samples in the training set, and thus they were not part of the training set.

As shown in Table~\ref{tab:newmal}, the detection rate on the test set is 71.96\%, which is quite satisfying if compared with the performances of other Android anti-malware solutions that are available in the market (See Section~\ref{sec:comparison}). Moreover, the false positive rate is around 7.52\%, which is acceptable if we consider that an individual user typically installs a few dozen applications, and thus it might receive a false alert occasionally. This casual alert allows the user that the application has some characteristics similar to badware, and so it can be used only if the source is trusted. It is also worth remarking that our classification of false positives is linked to the ground-truth classification provided by VirusTotal at the time of evaluation. It is not unlikely that some of these applications might turn out to be classified as malware by other anti-malware tools in the near future, as we have already noticed during the experiments. We expect in a future work to show how many applications were correctly predicted as being malicious before their signatures were created. However, our experience suggests that even if the application is benign but labeled as being potentially risky by \texttt{IntelliAV}, then the user might look for less risky alternatives applications in Google Play \cite{SecuRank}.
In fact, we believe that people should be aware of some applications that might be potentially harmful, even if it turns out not to be so, rather than missing some real threats.

\begin{table}[t]
\centering
\caption{Training on the set of samples described in Section~\ref{sec:data}, and testing on new samples in 2017. GT refers to the Ground-truth of samples.}\label{tab:newmal}
\begin{tabular}{c|c|cc}
\noalign{\smallskip}
\rowcolor{gray!60}
Train & \multicolumn{3}{c}{Test}\\
\multirow{2}{*}{\#Samples} & \multirow{2}{*}{GT (\#Samples)} & \multicolumn{2}{c}{Classified as}\\
 &  &  Malicious & Benign \\ \hline
\multirow{4}{*}{19,722} & Malicious (2311) & 1,663 & 648 \\
 & & (TPR = 71.96\%) \\
& Benign (2898) & 218 & 2,680 \\
 & & (FPR = 7.52\%) \\
\hline
\end{tabular}
\end{table}

\subsubsection{Challenging Modern AV vendors}
\label{sec:comparison}

There is a growth in the number of anti-malware vendors that
resort to machine learning approaches \cite{virustotal}. However, the foremost focus of these products appears to be on desktop malware, especially Windows PE malware. 
Based on the publicly available information, there are just a few pieces of evidence of two anti-malware vendors that use machine learning for Android malware detection, namely Symantec\cite{eweek} and TrustLook \cite{trustlook}, and their products are installed by more than 10 million users. 
Despite the lack of clearance of using machine learning by these tools to us,
we consider them as two candidates for comparison with \texttt{IntelliAV}. 
To provide a rational comparison, in addition to the Symantec and Trustlook products, we choose three other Android anti-malware products, i.e., AVG, Avast, and Qihoo 360, that are the most popular security tools among Android users as they have been installed more than 100 million times.\footnote{http://www.androidrank.org/}

\begin{figure}[t]
\pgfplotstableread{
0 1663
1 1580
2 1575
3 756
4 731
5 180
}\dataset
\centering
\resizebox{!}{7.0cm}
{
\begin{tikzpicture}
\begin{axis}[ybar,
        width=9cm,
        height=8cm,
        ymin=0,
        ymax=2311,        
        ylabel={Total test malware},
        xtick=data,
        xticklabels = {
            \strut \texttt{IntelliAV},
            \strut AV1,
            \strut AV2,
            \strut AV3,
            \strut AV4,
            \strut AV5
        },
        major x tick style = {opacity=0},
        minor x tick num = 1,
        minor tick length=2ex,
        every node near coord/.append style={
                anchor=west,
                rotate=90
        },
        ]
\addplot[draw=black,fill=blue!20, nodes near coords] table[x index=0,y index=1] \dataset; 
\end{axis}
\end{tikzpicture}
}
\caption{Comparison between the detection rate of IntelliAV with top five Android anti-malware. We didn't put the name of vendors as we don't aim to rank other anti-malware products.}
\label{fig:avsresults}

\end{figure}
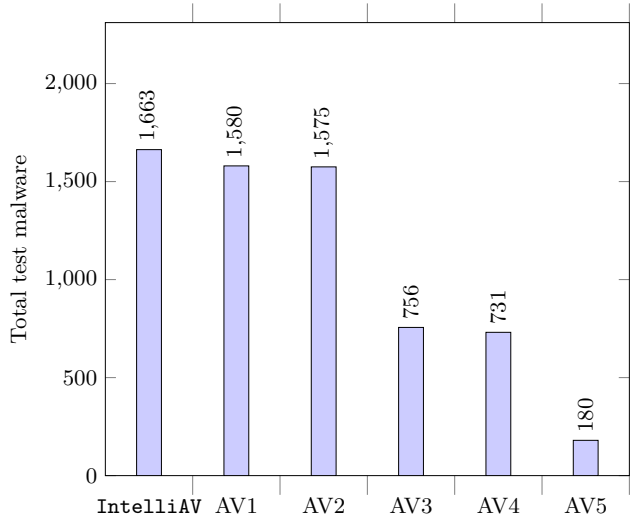

We compared the performances of \texttt{IntelliAV} on the test dataset (see Section~\ref{sec:unseenmalware}) with the ones attained by these five popular Android anti-malware.
As shown in Figure~\ref{fig:avsresults}, \texttt{IntelliAV} performs slightly better than two of the products used for comparison, while it outperforms the other three. As we gathered the label assigned by anti-malware products to the test samples at most two months after they are first seen in VirusTotal, the comparison could be more interesting if we had the label given to samples at the time they are first seen in the wild.

As a supplementary experiment, we carried out a measurement in detection performance by considering a set of top and very recent malware threats reported by four vendors, namely Check Point, Fortinet, Lookout, and Google (see Table~\ref{tab:com-trustlook}). The considerable performances of \texttt{IntelliAV} compared to the ones of other products, confirms the effectiveness of the selected lightweight features and the training procedure, especially if we consider that 21 of the analyzed samples were first seen before 2017, so it is expected that they can be detected by anti-malware tools either by signatures, or by the generalization capability provided by their machine learning engines.
If we have a close look at the two misclassified samples by \texttt{IntelliAV} (Table~\ref{tab:com-trustlook}), we can see that the associated risk scores are quite close to the decision threshold that we set at training time. The main reasons for the misclassification of these two samples can be related to the use of the \texttt{runtime.exec} API to run some shell commands, and to the presence of native-code, that is used to hide some of their malicious behaviors.

\begin{table}
\centering
\caption{Point to point comparison of \texttt{IntelliAV} and three anti-malware vendors on some recent and well-known malware reported by Check Point, Fortinet, Lookout, and Google from January to April of 2017. These samples were evaluated on an Android emulator. The \texttt{time} column refers to the required time for performing both feature extraction and classification on the emulator.}\label{tab:com-trustlook}
\resizebox{!}{6.8cm}{
\begin{tabular}{c|cc|ccc|ccc|c}
\hline\hline
\noalign{\smallskip}
\multicolumn{3}{c|}{} & \multicolumn{3}{c|}{\texttt{IntelliAV}} & \multicolumn{3}{c|}{2017 check} \\
\# & MD5 & Size & Unseen & time(s) & Risk Score & AV1 & AV3 & AV5 & VT 1st check \\ \hline
\rowcolor{gray!60}
& \multicolumn{9}{c}{Reported malware by Checkpoint \cite{checkpoint,checkpoint-gp,checkpoint-gp1,jsmalware}}\\
1 & {\tiny 60806c69e0f4643609dcdf127c8e7ef5} & 66 KB & \ding{51} & 0.38 & 83\% (\ding{51}) & (\ding{51}) & (\ding{51}) &(\ding{51}) & 2016-01 (02/56) \\
2 & {\tiny fcbb243294bb87b039f113352a8db158} & 12.4 MB & \ding{51} & 0.40 & 37\% (\ding{55}) & (\ding{51}) &(\ding{55}) &(\ding{55})  & 2016-03 (19/55)  \\
3 & {\tiny 4e91ff9ac7e3e349b5b9fe36fb505cb4} & 48 KB & \ding{51} & 0.37 & 93\% (\ding{51}) & (\ding{51}) &(\ding{51}) &(\ding{51}) & 2016-03 (13/57)  \\
4 & {\tiny 944850ee0b7fc774c055a2233478bb0f} & 842 KB & \ding{51} & 0.51 & 98\% (\ding{51}) & (\ding{55}) &(\ding{55}) &(\ding{55}) & 2014-02 (00/48)  \\
5 & {\tiny 629da296cba945662e436bbe10a5cdaa} & 3.7 MB & \ding{51} & 0.69 & 92\% (\ding{51}) & (\ding{51}) & (\ding{55})&(\ding{51}) & 2014-07 (13/51)  \\
6 & {\tiny 1aac52b7d55f4c1c03c85ed067bf69d9} & 3.5 MB & \ding{51} & 0.75 & 94\% (\ding{51}) & (\ding{51}) &(\ding{51}) &(\ding{51}) & 2013-11 (23/47)  \\
7 & {\tiny 379ec59048488fdb74376c4ffa00d1be} & 2.2 MB & \ding{51} & 0.57 & 79\% (\ding{51}) & (\ding{51}) &(\ding{51}) &(\ding{51}) & 2015-09 (26/56)  \\
8 & {\tiny d5f5480a7b29ffd51c718b63d1ffa165} & 9.1 MB & \ding{51} & 0.82 & 89\% (\ding{51}) & (\ding{51}) & (\ding{55}) &(\ding{55}) & 2015-12 (03/55)  \\
9 & {\tiny 4d904a24f8f4c52726eb340b329731dd} & 13.2 MB & \ding{51} & 0.95 & 72\% (\ding{51}) & (\ding{51}) & (\ding{55}) &(\ding{55}) & 2014-08 (11/51)  \\
10 & {\tiny 59b62f8bc982b31d5e0411c74dbe0897} & 2.5 MB & \ding{51} & 0.45 & 83\% (\ding{51}) & (\ding{51}) &(\ding{51}) &(\ding{51}) & 2016-01 (31/55)  \\
11 & {\tiny 9ed38abb335f0101f55ad20bde8468dc} & 8.1 MB & \ding{51} & 0.77 & 67\% (\ding{51}) & (\ding{51}) & (\ding{55}) &(\ding{55}) & 2016-02 (16/55)  \\
12 & {\tiny 4a3a7b03c0d0460ed8c5beff5c20683c} & 575 KB & \ding{51} & 0.42 & 68\% (\ding{51}) & (\ding{51}) &(\ding{51}) &(\ding{51}) & 2017-03 (00/55)  \\
13 & {\tiny 660638f5212ef61891090200c354a6d5} & 32.7 MB & \ding{51} & 1.13 & 96\% (\ding{51}) & (\ding{51}) & (\ding{55}) &(\ding{51}) & 2016-07 (13/55)  \\
14 & {\tiny f48122e9f4333ba3bb77fac869043420} & 349 KB & \ding{51} &  0.40 & 81\% (\ding{51}) & (\ding{51}) & (\ding{55}) &(\ding{51}) & 2015-09 (04/57)  \\
15 & {\tiny 0e987ba8da76f93e8e541150d08e2045} & 12.8 MB & \ding{51} &  0.98 & 88\% (\ding{51}) & (\ding{51}) & (\ding{55}) &(\ding{55}) & 2017-03 (07/60)  \\
16 & {\tiny 51c328fccf1a8b4925054136ccdb1cda} & 874 KB & \ding{51} & 0.44 & 83\% (\ding{51}) & (\ding{55}) &(\ding{51}) &(\ding{55}) & 2014-08 (05/53)  \\
17 & {\tiny 3f188b9aa8f739ee0ed572992a21b118} & 1.57 MB & \ding{51} &  0.48 & 89\% (\ding{51}) & (\ding{51}) &(\ding{51}) &(\ding{51}) & 2014-04 (24/51)  \\
18 & {\tiny 7fff1e78089eb387b6adfa595385b2c9} & 13.4 MB & \ding{51} &  0.52 & 63\% (\ding{51}) & (\ding{51}) &(\ding{51}) &(\ding{51}) & 2015-03 (02/57)  \\
\hline
19 & {\tiny 2b83bd1d97eb911e9d53765edb5ea79e} & 2.3 MB & \ding{51} & 0.43 & 77\% (\ding{51}) & (\ding{55}) &(\ding{51}) &(\ding{55}) & 2017-01 (16/58)  \\
\hline
20 & {\tiny 48ff097022ea7886b53f80edf2972033} & 1.3 MB & \ding{51} & 0.47 & 63\% (\ding{51}) & (\ding{55}) &(\ding{51}) &(\ding{55}) & 2017-03 (28/59)  \\
21 & {\tiny a3836485ecac78f576e1753269350824} & 14.6 MB & \ding{51} & 0.84 & 38\% (\ding{55}) & (\ding{55}) &(\ding{55}) &(\ding{55}) & 2016-12 (14/57)  \\
22 & {\tiny a4e75471dbf0bb0d3ec26d854cb7fe12} & 14.1 MB & \ding{51} & 0.72 & 62\% (\ding{51}) & (\ding{55}) &(\ding{51}) &(\ding{55}) & 2016-12 (10/56)  \\
23 & {\tiny 7253e0a13d2d1db1547e9984a4ce7abd} & 1.3 MB & \ding{51} & 0.57 & 63\% (\ding{51}) & (\ding{55}) &(\ding{55}) &(\ding{55}) & 2017-03 (26/59)  \\
\hline
24 & {\tiny 84a62599a40e36be2180485245e8123f} & 5.7 MB & \ding{51} & 0.17 & 81\% (\ding{51}) & (\ding{55}) &(\ding{55}) &(\ding{55}) & 2017-05 (02/62)  \\
25 & {\tiny c2132651331a77d41a323fefa71bfbd0} & 5.1 MB & \ding{51} & 0.29 & 87\% (\ding{51}) & (\ding{55}) &(\ding{55}) &(\ding{55}) & 2017-05 (04/59)  \\
\hline
\rowcolor{gray!60}
& \multicolumn{9}{c}{Reported malware by Fortinet \cite{fortinet1,fortinet2,fortinet3,fortinet4}}\\
26 & {\tiny 193058ae838161ee4735a9172ebc25ec} & 1.4 MB & \ding{51} &  0.56 & 89\% (\ding{51}) & (\ding{51}) &(\ding{55}) &(\ding{55}) & 2017-01 (05/24)  \\ \hline
27 & {\tiny f479f2a29354a8b889cb529a2ee2c1b4} & 1.1 MB & \ding{51} & 0.35 & 61\% (\ding{51}) & (\ding{55}) &(\ding{51}) &(\ding{55}) & 2017-03 (12/59)  \\ \hline
28 & {\tiny cad94ac28640c771b1d2de5e786dc352} & 776 KB & \ding{51} &  0.37 & 96\% (\ding{51}) & (\ding{51}) &(\ding{51}) &(\ding{55}) & 2016-11 (20/56)  \\ \hline
29 & {\tiny 40507254b8156de817f02c0ed111e99f} & 0.2 MB & \ding{51} &  0.37 & 83\% (\ding{51}) & (\ding{51}) &(\ding{51}) &(\ding{51}) & 2016-11 (08/57)  \\ \hline
\rowcolor{gray!60}
& \multicolumn{9}{c}{Reported malware by Lookout and Google \cite{pegasus-lookout,pegasus-google}}\\
30 & {\tiny cc9517aafb58279091ac17533293edc1} & 57 KB & \ding{51} &  0.63 & 89\% (\ding{51}) & (\ding{55}) &(\ding{55}) &(\ding{55}) & 2016-02 (00/53)  \\ 
31 & {\tiny 7c3ad8fec33465fed6563bbfabb5b13d} & 252 KB & \ding{51} &  0.37 & 82\% (\ding{51}) & (\ding{55}) &(\ding{51}) & (\ding{55}) & 2017-04 (03/60)  \\ 
32 & {\tiny 3a69bfbe5bc83c4df938177e05cd7c7c} & 19 KB & \ding{51} &  0.36 & 81\% (\ding{51}) & (\ding{55}) &(\ding{55}) &(\ding{55}) & 2017-04 (01/60)  \\ 
\hline\hline
\noalign{\smallskip}
\multicolumn{6}{r|}{{\LARGE $\frac{30}{32}$}~\Smiley[2]} & {\LARGE $\frac{19}{32}$} & {\LARGE $\frac{16}{32}$} & {\LARGE $\frac{12}{32}$}\\
\end{tabular}
}
\end{table}

\subsubsection{Independent Test by a Third Party}
\label{sec:AV-TEST}
\begin{figure}[t]
\pgfplotstableread{
0 96
1 100
2 100
3 100
4 100
5 100
6 100
7 88
8 100
9 100
10 98
11 70
12 100
13 89
14 92
}\dataset
\centering
\resizebox{!}{8.0cm}
{
\begin{tikzpicture}
\begin{axis}[ybar,
        width=16cm,
        height=9cm,
        ymin=0,
        ymax=100,        
        ylabel={Detection Rate (\%)},
        xtick=data,
        xticklabels = {
            \strut \textbf{{\Large Sum Total}} (478/500),
            \strut BD/DroidKungFu (15/15),
            \strut BD/Other (31/31),
            \strut TR-SMS/Fakeinst (31/31),
            \strut TR-SMS/Other (13/13),
            \strut TR-SMS/Smsthief (23/23),
            \strut TR-Spy/Androrat (22/22),
            \strut TR-Spy/Other (36/41),
            \strut TR-Spy/Smforw (45/45),
            \strut TR-Spy/Smsspy (51/51),
            \strut TR/Banker (78/80),
            \strut TR/Fakeapp (7/10),
            \strut TR/Hqwar (20/20),
            \strut TR/Locker (39/44),
            \strut TR/Other (83/90),
        },
        major x tick style = {opacity=0},
        x tick label style={rotate=90},
        minor x tick num = 1,
        minor tick length=1ex,
        every node near coord/.append style={
                anchor=west,
                rotate=90
        },
        ]
\addplot[draw=black,fill=red!20, nodes near coords] table[x index=0,y index=1] \dataset; 
\end{axis}
\end{tikzpicture}
}
\caption{Results of the independent test on IntelliAV in July 2017. BD and TR refer to backdoor and trojan respectively. The detection rate is separated based on the malware families as well.}
\label{fig:avtest}

\end{figure}
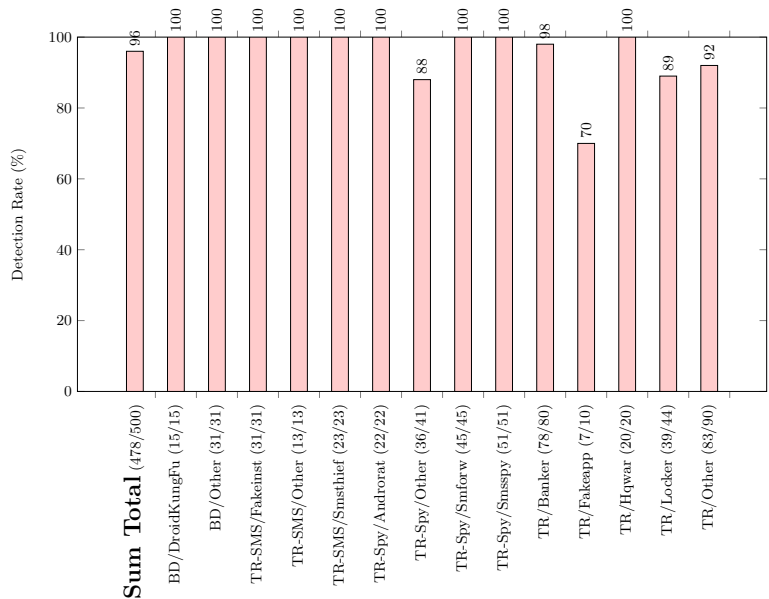

To avoid any bias regarding testing \texttt{IntelliAV}, we requested a third-party independent anti-malware testing organization to assess our tool. The test performed on a test suit involving 500 most common and recent Android malware, which all have been first seen by VirusTotal in 2017 ---more recent than the samples in our training test. Interestingly, as is shown in Figure~\ref{fig:avtest}, \texttt{IntelliAV} can successfully classify 478 of the samples as malware, i.e., 96\% malware detection rate. Except for \texttt{Fakeapp} family, \texttt{IntelliAV} achieves very good results on the rest of malware families. 
As specimens in \texttt{Fakeapp} family regularly abuse legitimate application by repackaging them, the presence of more benign characteristics in such malware seems to be the main reason of misclassification.

To measure the false positive rate, \texttt{IntelliAV} was tested on 
50 benign applications that have been downloaded more than 10,000 times from Google Play, as well as were not tagged as malware by any anti-malware product. Two out of the 50 apps were misclassified by \texttt{IntelliAV}. However, one of the two misclassified samples has been removed from Google Play recently. Although the main reason for elimination is not clear, the application surely violates some of the Google Play policies and \texttt{IntelliAV} seems to detect such a kind of violation perfectly.

\subsection{Stop Droppers on Device}
\label{sec:droppers}

Dynamic code loading techniques are employed by both the benign and malicious applications \cite{poeplau14:executethis,Zhauniarovich:2015:SAP}. There are different techniques to dynamically load a piece of code in Android applications, and the technique that turns to be more popular among malware developers has not been yet clearly evaluated. However, there are many reports by anti-malware vendors \cite{Trendmicro-dropper,Eset-dropper} that designates APK installation technique, also known as droppers, as one of the most popular approaches by which attackers deliver a malware. 
A dropper is a malware installer that surreptitiously carries any kind of malicious software so they can be executed on the compromised machine. 
They do not carry any malicious activities by themselves, but they just open a way for the attack by downloading/decompressing and then installing the core malware payload onto a target machine without detection.
Depending on the way they retrieve the payload, i.e., by downloading or decompressing, they might be called \texttt{Downloader Trojans} or \texttt{Wrappers}.

\tikzstyle{block} = [rectangle, draw, fill=gray!20, 
    text width=5em, text centered, rounded corners, minimum height=4em]
    \tikzstyle{recblock} = [rectangle, draw, fill=gray!40, 
    text width=5em, text centered, minimum height=2em]
    \tikzstyle{subblock} = [rectangle, draw, fill=gray!9, 
    text width=5.5em, text centered, rounded corners, minimum height=2em]
\tikzstyle{dashedblock} = [rectangle, draw, dashed, fill=gray!9, 
    text width=5.5em, text centered, rounded corners, minimum height=2em]
\tikzstyle{line} = [draw, -latex', line width=4mm]
    \tikzstyle{io} = [rectangle, 
    text width=5em, text centered, rounded corners, minimum height=2em]
        \tikzstyle{nio} = [rectangle,
    text width=5em, text centered, rounded corners, minimum height=1em]
        \tikzstyle{nblock} = [rectangle, draw, fill=gray!20,
    text width=5em, text centered, rounded corners, minimum height=4em]
    \tikzstyle{sblock} = [rectangle, draw, fill=gray!20,
    text width=5em, text centered, rounded corners, minimum height=2em]
\tikzstyle{ndashedblock} = [rectangle, draw, dashed, fill=white!9, 
    text width=5.5em, text centered, rounded corners, minimum height=2em]
\tikzstyle{horizontal-dotted} = [dash pattern = on 1pt off 4pt,
      line width =1.4pt]

\begin{figure}[t]
\begin{center}
\resizebox{!}{8.5cm}{
\begin{tikzpicture}[node distance = 4cm, auto]
    \node[inner sep=0pt, xshift=-5cm] (dropper)
    {
    \includegraphics[width=0.35\textwidth]{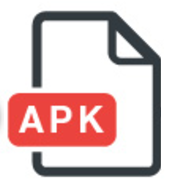}
    };
    \node[inner sep=0pt, right of=dropper, xshift=23cm] (malicious)
    {
    \includegraphics[width=.4\textwidth]{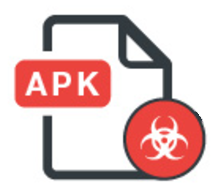}
    };
    \node[inner sep=0pt, below of=dropper, xshift=0cm, yshift=-13cm] (avdropper)
    {
    \includegraphics[width=.9\textwidth]{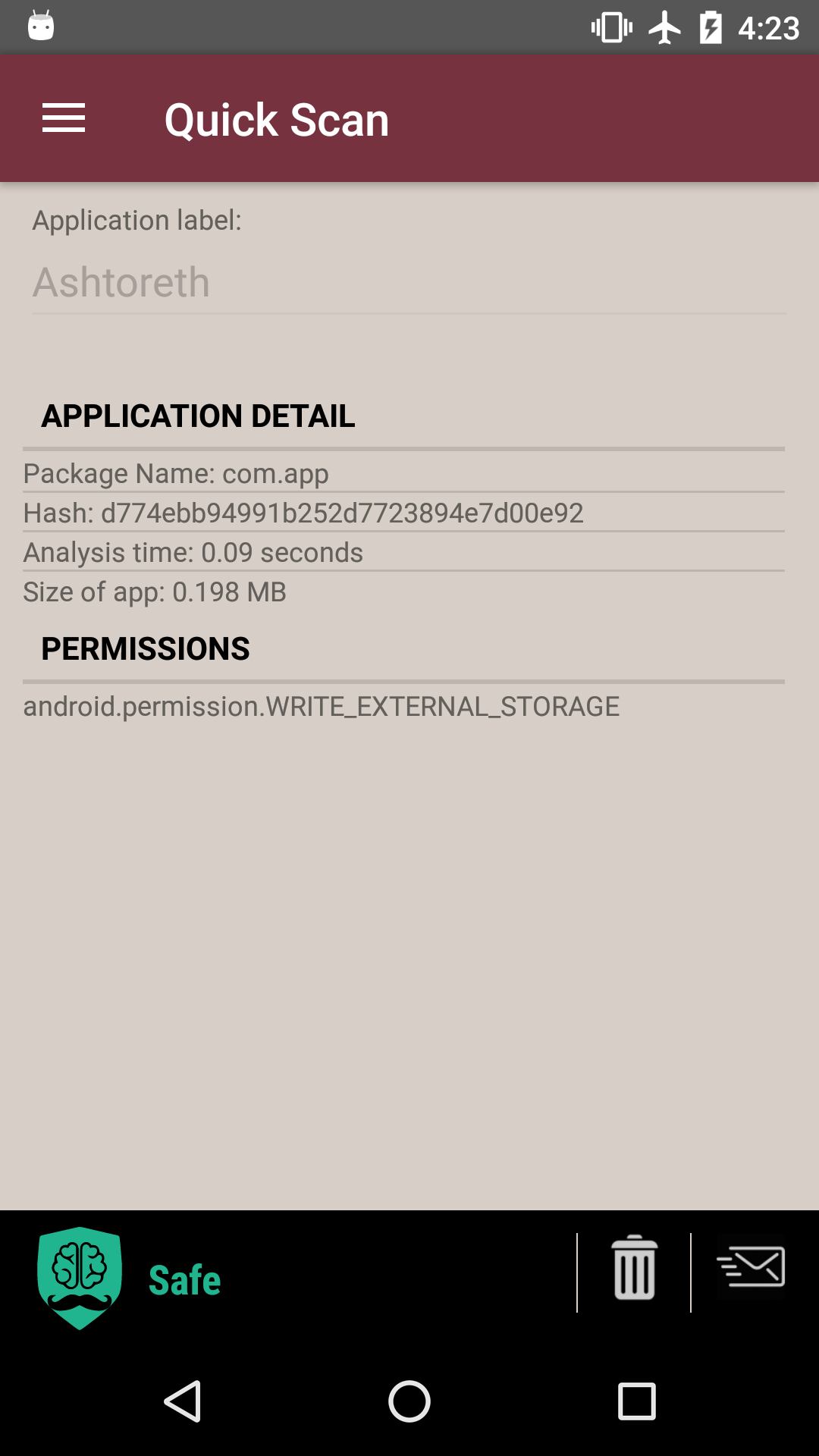}
    };
    \node[inner sep=0pt, right of=avdropper, xshift=10cm] (drop)
    {
    \includegraphics[width=.9\textwidth]{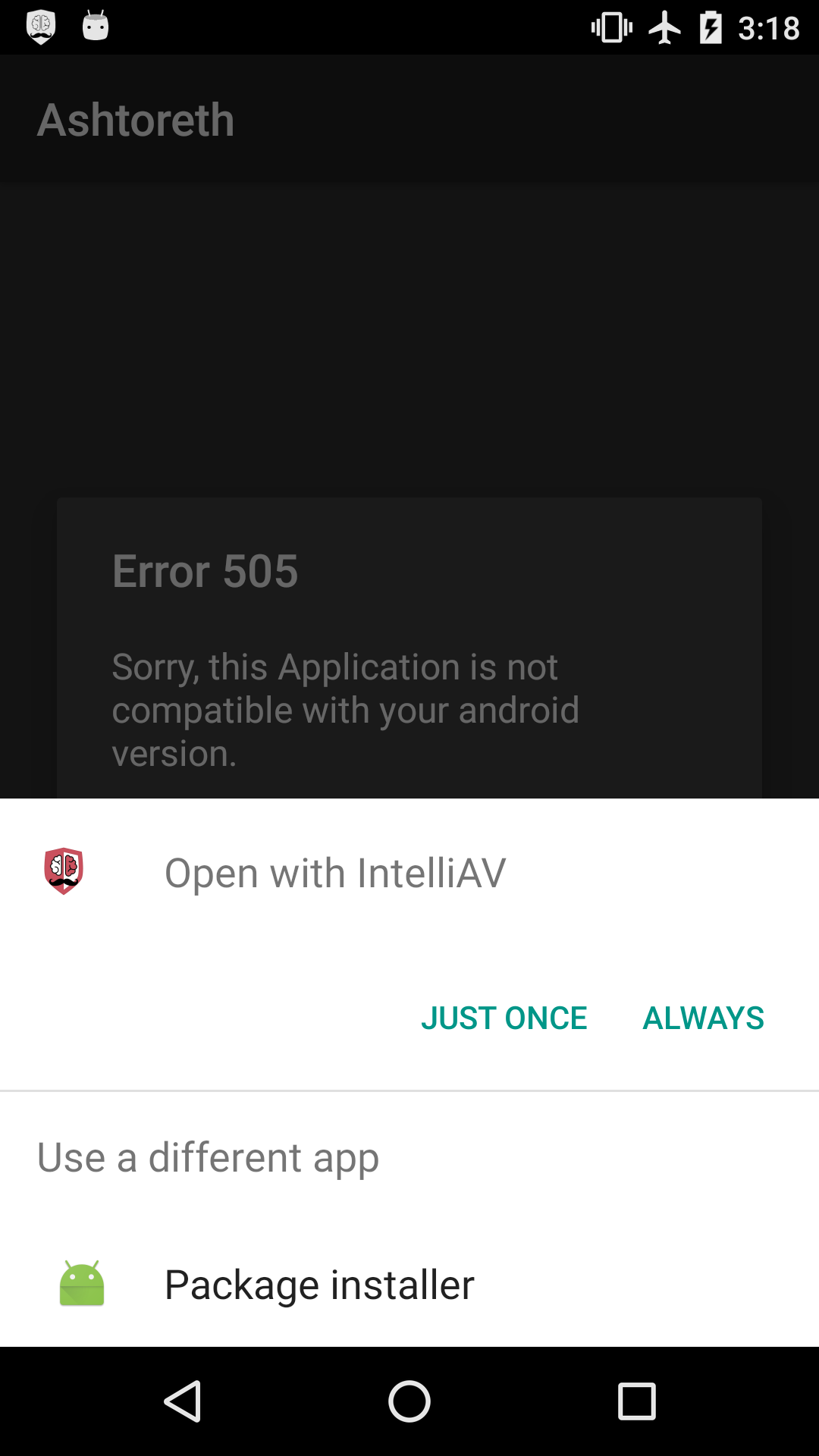}
    };
    \node[inner sep=0pt, right of=drop, xshift=10cm] (avmalicious)
    {
    \includegraphics[width=.9\textwidth]{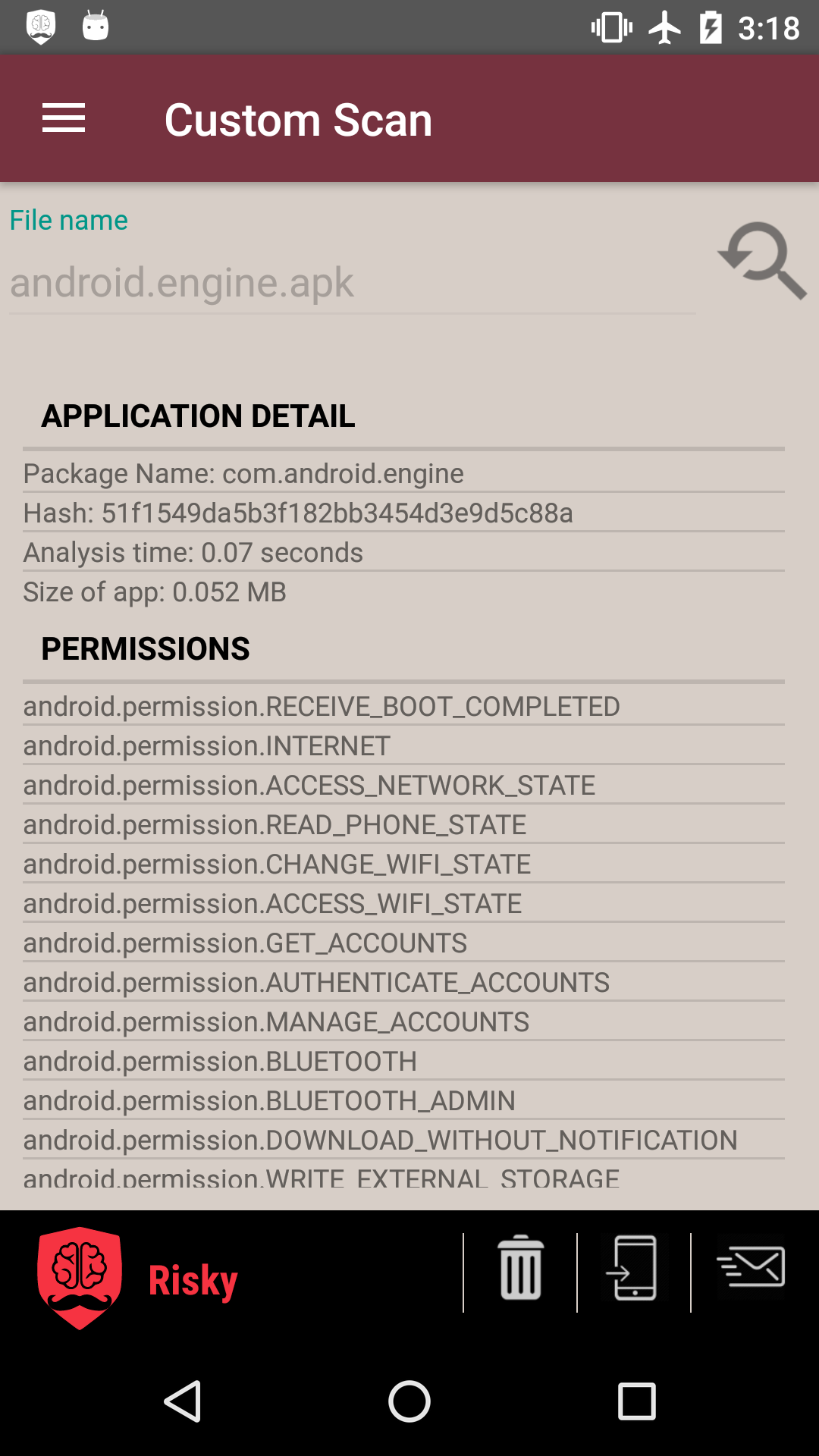}
    };

     \draw[horizontal-dotted, right of=avmalicious, yshift=-15em, xshift=0cm](-15,1)--(25,1);
     \node [io, yshift=0em, below of=dropper, text width = 2cm, yshift=2em, xshift=-3em] (ows) {\uline{{\huge \textbf{Dropper APK}}}};
     \node [io, yshift=0em, below of=malicious, text width = 2cm, yshift=2em, xshift=-4em] (ows) {\uline{{\huge \textbf{Dropped APK}}}};
    \path [line,dashed] (avdropper) -- (drop);
    \path [line,dashed] (drop) -- (avmalicious);
    \path [line] (dropper) -- node[xshift=0cm, yshift=0em, text width=20cm, align=center]{{\huge \textbf{Drop}} \\ {\huge Decode the wrapped payload from a resource/asset inside the dropper} \\ {\huge Download the payload from Internet} }(malicious);

\end{tikzpicture}
}
\end{center}
\caption{Detecting dropper attacks by IntelliAV on the device. Offline machine learning techniques would fail to detect this kind of attack. The example is a very recent Android malware, which was reported in July 2017 by Trendmicro \cite{Trendmicro-dropper}}\label{fig:dropper}
\end{figure} 

The overall technique employed by droppers is summarized in Figure~\ref{fig:dropper}. The upper part of the figure is the overall scheme of the attack technique, and the bottom part is showing how \texttt{IntelliAV} can stop this attack on the device when the malware has to reveal its full capabilities while offline machine learning techniques would fail. To dissect this attack vector in more detail, we evaluate a very recent dropper Android malware, which was reported in July 2017 by Trendmicro \cite{Trendmicro-dropper}. 
This malware misleads users to install a secondary app by showing a fake message.
When the dropper app\footnote{Dropper APK MD5: d774ebb94991b252d7723894e7d00e92} is launched, an error pops out that: ``\textit{Sorry, the application is not compatible with your android version}". Meanwhile, it decodes a string from the APK resource file, which is actually a malicious APK\footnote{Dropped APK MD5: 51f1549da5b3f182bb3454d3e9d5c88a}, and requests the user to install it.
As the dropper does not contain malicious activities and just needs one permission to drop the secondary APK, \texttt{IntelliAV} does not produce any risk alert. However, before the user is infected by the dropped APK, they can request \texttt{IntelliAV} to scan it before installation (i.e., by its custom scan capability), which results in the detection of the real malicious app.

It is obvious that \texttt{IntelliAV} can generalize this detection and prevention approach to any malware that uses APK installation technique. Another example is an Android malware\footnote{Dropper APK MD5: 1f41ba0781d51751971ee705dfa307d2}, reported by an ESET researcher \cite{Eset-dropper}, that asks just for the storage permission to save the the secondary loaded malicious APK\footnote{Dropped APK MD5: 90886b56372f5191a78a20dcb3f9fe6e}. After the first APK executes, it immediately drops the secondary APK, embedded in an image located in an APK asset, and requests the user to install it. In another report by the same vendor, a dropper\footnote{Dropper APK MD5: add7572e94b54bb408d34c8c19972b0c} is a kind of logic bomb \cite{androidlb} and drops a malicious apk\footnote{Dropped APK MD5: 243028fe2a1a9358187d6e694343fcd9} twenty minutes after the first run \cite{Eset-dropper-gp}.

More advanced offline machine learning techniques relying on static analysis, even those deeply look inside an APK to find an embedded malware like DroidSieve \cite{DroidSieve}, cannot model all of the feasible dropping techniques because the secondary app can be concealed by custom methods. Although offline approaches relying on dynamic analysis \cite{copperdroid} might be able to detect this kind of attack if they are system-centric and designed perfectly to disclose logic bombs, there has not been any research specifically to address this issue.

\subsection{Robustness against Obfuscation}
\label{sec:obfuscation}
Obfuscation techniques \cite{Rastogi:2013:DEA,vb14}
are widely used by both the benign and malicious Android applications to protect themselves against reverse engineering. Although it is clear that there is no bulletproof approach, a formidable detection system should not be influenced by obfuscation dramatically. Hence, to evaluate the resiliency of \texttt{IntelliAV}, we employ a recently released obfuscation framework, namely AVPASS \cite{jung:avpass-bh}, that has been shown to be able to evade almost all of the commercial Android anti-malware products.
AVPASS uses different common obfuscation techniques to encrypt and remove/add features from/to an application. These techniques are summarized in Table~\ref{tab:obfuscation}. 

\begin{table}[]
\centering
\caption{Obfuscation techniques used for evaluating IntelliAV. These techniques preserve the original functionality of the application. The techniques like removing permissions are not considered here as they destroy the functionality of the application.}
\label{tab:obfuscation}
\begin{tabular}{clc} \\
 \hline
& \multicolumn{1}{c}{ \textbf{Obfuscation Type}} & Affect on IntelliAV   \\
\hline
1 & Insert dummy bytes between all instructions   & \ding{55}    \\
2 & Insert benign permissions to AnadroidManifest.xml & Possible \\
3 & Image file obfuscation    &    \ding{55}                    \\
4 & Change variables name   &      \ding{55}                    \\
5 & Modify user-defined class/package name  &     \ding{55}     \\
6 & Insert API between two existing APIs  &    Possible       \\
7 & Encrypt strings &  \ding{55} \\
8 & API hiding by Java Reflection &  Possible \\
9 & Resource XML obfuscation  &   \ding{55} \\
\hline
\end{tabular}
\end{table}

Among the aforementioned obfuscation techniques, three of them might directly affect on \texttt{IntelliAV}, and they are 1) adding benign permissions, 2) hiding APIs by Java reflection, and 3) inserting APIs. To realize how much the techniques in Table~\ref{tab:obfuscation} can undermine \texttt{IntelliAV}, we randomly select ten small size Android malware (see Table~\ref{tab:obfuscation}). The main reason of being small size is that employing all of the obfuscation techniques needs a huge processing time and large sized apps cannot be done on a laptop. 

We investigate \texttt{IntelliAV} in two different scenarios, i.e., testing on the ten original malware as well as testing on the obfuscated variant of the malware. Then, we use the identical approach for testing the samples against those anti-malware products that are accessible in Virus Total. 
Interestingly, \texttt{IntelliAV} is completely robust against all of the obfuscation techniques in Table~\ref{tab:obfuscation}. While the obfuscated variant can simply evade most of the anti-viruses, evasion from \texttt{IntelliAV} needs more elaborate efforts. Among the anti-malware that have detected the obfuscated variant,
three of them performs very well. However, it seems that they fingerprint AVPASS footsteps because they detect even a benign app, obfuscated by AVPASS, as a malware. Moreover, they use the same family name from all the obfuscated malware by AVPASS.
In this case, if an attacker applies a custom obfuscation/evasion technique, they can readily dodge not carefully crafted anti-malware products. It is worth to mention that none of the antimalware, which have claimed they are based on machine learning, are able to detect more than five samples. This evaluation again arises the attention toward the lack of either use of machine learning or generalization of the detection model.

\begin{table}
\centering
\caption{Point to point comparison of \texttt{IntelliAV} and VirusTotal on some randomly selected malware as well as obfuscated variant of the malware. The percentage numbers show the risk score calculated by IntelliAV. The \texttt{VT} column shows the number of anti-malware vendors that are able to detect a malware before and after obfuscation. Almost all of the anti-malware products have difficulty to detect the obfuscated malware. There are three anti-malware that able to detect most of obfuscated malware. However, it seems that they fingerprint AVPASS as they detect even obfuscated benign apps by AVPASS as malware. The test has been performed on VT in early August, 2017.}
\resizebox{!}{3.0cm}{
\begin{tabular}{c|c|cc|cc}
\hline\hline
\noalign{\smallskip}
\multicolumn{2}{c|}{} & \multicolumn{2}{c|}{\texttt{Original Malware}} & \multicolumn{2}{c}{Obfuscated Variant} \\
\# & MD5 & IntelliAV & VT & IntelliAV & VT \\ \hline
1 & {\tiny 60806c69e0f4643609dcdf127c8e7ef5} & 82\% (\ding{51}) & 31/59 & 84\% (\ding{51}) & 4/60 \\
2 & {\tiny 4e91ff9ac7e3e349b5b9fe36fb505cb4} & 87\% (\ding{51}) & 34/60 & 75\% (\ding{51}) & 4/61 \\
3 & {\tiny 26b8a840b4cc15d0b05533705d21854b} & 100\% (\ding{51}) & 43/60 & 98\% (\ding{51}) & 3/61 \\
4 & {\tiny 19723d489712cff0ea2907f142937b8b} & 100\% (\ding{51}) & 40/62 & 84\% (\ding{51}) & 5/61 \\
5 & {\tiny 4760ac11bf995d26c0486cf3b73f5f9a} & 100\% (\ding{51}) & 43/61 & 92\% (\ding{51}) & 6/60 \\
6 & {\tiny 0085ae9415d115a6bde1e9ff72b6dc7f} & 100\% (\ding{51}) & 43/59 & 98\% (\ding{51}) & 3/60 \\
7 & {\tiny 9eebbc312fa1cdc891e151f367e6bd9d} & 84\% (\ding{51}) & 31/61 & 81\% (\ding{51}) & 8/59 \\
8 & {\tiny 711d00941aa78e912051d644f84748be} & 100\% (\ding{51}) & 31/62 & 98\% (\ding{51}) & 7/61 \\
9 & {\tiny cabdc04298dbf6bce3857328d24c96dd} & 100\% (\ding{51}) & 38/61 & 84\% (\ding{51}) & 6/58 \\
10 & {\tiny a0113b794e69d7ca1d603aadf977b8a7} & 100\% (\ding{51}) & 46/60 & 92\% (\ding{51}) & 7/61 \\
\hline\hline
\noalign{\smallskip}
\end{tabular}
}
\end{table}

\subsection{\texttt{IntelliAV} Overhead on Device}
\label{sec:overhead}

Despite the belief that running a detection model on mobile devices is computationally infeasible, we acknowledge the efficiency of \texttt{IntelliAV} by exposing the time consumption for the feature extraction as well as the classification phases. We select some popular medium/large-sized applications and analyze them by \texttt{IntelliAV} on three devices with different technical specifications. The three mobile devices used for the reported experiments are a Samsung Galaxy S6 Edge (released in April, 2015), a Huawei P8 Lite (released in May, 2015), and an LG D280 L65 (released in June, 2014), which respectively have 3GB, 2GB, and 1GB of RAM.
In addition, we computed the time required on the Android Emulator that is dispatched along with Android Studio.
The time is simply computed by specifying a timer before starting the feature extraction procedure, that stops when the features from both the manifest and the dex code are extracted. For classification, the reported time refers to the interval between the time in which the feature vector is passed to the model, and the time of production of the risk score.
The time required to load the model is negligible, and so we are not reporting it for the sake of clarity.

As shown in Table~\ref{tab:Overhead}, the time required to analyze even large applications 
is less than 10 seconds, which makes \texttt{IntelliAV} practical and reasonable as the number of installed applications on each device is not too large, and the computational power of mobile devices is increasing even on cheap models. The classification part is performed in native code, that provides a fast execution. As expected, it can be noted that the largest fraction of the time required by \texttt{IntelliAV} is spent for feature extraction, especially for the extraction of the API features. 
Extraction of API features is even much slower in the case an application is made up of multiple dex files. For instance, the Uber app is made up of 10 dex files, so that searching for a specific API requires much more time compared to applications having just one dex file.

\begin{table}[t]
\centering
\caption{Overhead of \texttt{IntelliAV} on different devices for very large applications. F.E. refers to feature extraction time and C. refers to classification time. The number in parenthesis shows the RAM size of the device.}\label{tab:Overhead}
\resizebox{!}{2.6cm}{
\begin{tabular}{c|c|cc|cc|cc|cc}
\hline
\noalign{\smallskip}
 & & \multicolumn{2}{c}{Galaxy S6 Edge} & \multicolumn{2}{|c|}{Huawei P8 Lite} & \multicolumn{2}{c}{LG D280 L65} & \multicolumn{2}{|c}{Emulator} \\
 &  & \multicolumn{2}{c}{Marshmallow (3GB)} & \multicolumn{2}{|c|}{Lollipop (2GB)} & \multicolumn{2}{c}{KitKat (1GB)} & \multicolumn{2}{|c}{Marshmallow (1.5GB)} \\
App & APK Size (MB) & F.E. (s) & C. (s) & F.E. (s) & C. (s) & F.E. (s) & C. (s) & F.E. (s) & C. (s) \\ \hline
Google Trips 		& 8.19 	& 0.67 		& 0.003 & 0.82 & 0.005 & 3.86 & 0.012 & 0.43 & 0.001 \\
LinkedIn Pulse 		& 12.9 	&  1.28		& 0.003 & 1.14 & 0.005 & 4.40 & 0.012 & 0.55 & 0.001 \\
Stack Exchange 	& 8.15 	& 1.27 		& 0.004 & 1.27 & 0.006 & 5.13 & 0.014 & 0.60 & 0.001 \\
Telegram 				& 12.41 & 1.36 	& 0.005 & 1.74 & 0.007 & 5.52 & 0.016 & 0.69 & 0.002 \\
WhatsApp 			& 27.97 & 2.29 	& 0.006 & 3.22 & 0.008 & 12.91 & 0.018 & 1.10 & 0.002  \\
SoundCloud 			& 33.14 & 2.67	& 0.006 & 2.84 & 0.008 & 11.83 & 0.018 & 1.14 & 0.002  \\
Spotify 				& 34.65 & 2.51 	& 0.006 & 3.03 & 0.008 & 13.67 & 0.018 & 1.22 & 0.002 \\
Twitter 				& 31.77 & 4.53 	& 0.004 & 5.95 & 0.006 & 24.46 & 0.016 & 2.26 & 0.002 \\
LinkedIn 				& 40.39 & 4.67 	& 0.004 & 4.69 & 0.006 & 16.73 & 0.016 & 2.40 & 0.001 \\
Airbnb 					& 54.34 & 8.24 	& 0.006 & 8.79 & 0.008 & 35.71 & 0.018 & 4.23 & 0.002 \\
Messenger 			& 59.43 & 5.85 	& 0.011 & 7.94 & 0.013 & 19.13 & 0.028 & 3.35 & 0.004 \\
Uber 					& 37.26 & 6.66 	& 0.004 & 7.64 & 0.006 & 43.88 & 0.016 & 4.29 & 0.002 \\
\hline
\hline
\noalign{\smallskip}
Average 				& 30.05 & 3.50 	& 0.005 & 4.08& 0.007 & 16.43 & 0.016 & 1.86 & 0.002 \\
\hline
\end{tabular}
}
\end{table}
\section{Limitations}
\label{sec:limitations}
As far as \texttt{IntelliAV} is based on static analysis, it inherits some of the well-known limitations of static analysis approaches. For instance, although we partially addressed dynamic code loading techniques, the more complicated one, which hides the malicious behavior in the native code, might affect our system. Moreover, in the current proposed implementation, \texttt{IntelliAV} cannot detect the malicious actions executed by JavaScript.
In addition, we are aware that the system can be a victim of evasion techniques against the learning approach, such as 
mimicry attacks that let an attacker inject some data to the app so that its features resemble the ones of benign apps \cite{Demontis:2017,acsac:2017,Grosse2017}.
Consequently, more methodological and experimental analysis will be needed to make a quantitative evaluation of the robustness of \texttt{IntelliAV} in an adversarial environment, to provide the system with the required hardening. 
Nonetheless, we believe that the good performance of the proposed system against a few obfuscation techniques (e.g., adding benign permissions/APIs as well as hiding APIs) is a good starting point for further development. Moreover, employing the multiple classifier systems approaches, considering a larger number of semantic features, as well as performing a fine-grained classifier parameter tuning, can provide a degree of robustness against adversarial attacks against the machine learning engine.
\section{Conclusions and future work} \label{sect:conclusions}
In this work, we investigated the practicality of building a learning-based anti-malware tool for the devices running Android platform. To consider both the effectiveness and the efficiency of the tool, we emphasize on a careful selection of a set of lightweight features, as well as a solid training phase.
The reported results show that \texttt{IntelliAV} is robust against common obfuscation techniques. In addition, as far as \texttt{IntelliAV} runs on the device, it can track and scan all downloaded, dropped, and installed apps on the fly, which makes it more robust compared to off-device systems.
Our tool will be freely available so that it can help the end user to provide easy protection on the device, as well as allowing researchers to better explore the idea of having intelligent security systems on mobile devices.
As a future plan, we aim to address the limitations of \texttt{IntelliAV}, to improve its robustness against attacks on the machine learning engine, while keeping the efficiency intact.

\section{Acknowledgment}
This work has been supported by the project ``PISDAS'' jointly carried out with Innovery SpA, funded by the Regional Administration of Sardinia under the program ``Pacchetti Integrati di Agevolazione Industria Artigianato e Servizi ai sensi della deliberazione G.R.  n. 46 del  31/10/2013'', proposal 165. We appreciate VirusTotal's collaboration for providing us the access to a large set of Android applications. As a note, some of the antivirus vendors existing in VirusTotal platform do not always exactly perform the same as their public commercial versions. However, it is the best available option that can be used for a large scale testing. Moreover, we would like to acknowledge the third party antivirus testing organization for evaluating \texttt{IntelliAV} for free.


\bibliography{IntelliAV}

\end{document}